\documentclass[%
 aip,
 jcp,
 amsmath,amssymb,
reprint,%
]{revtex4-2}

\usepackage{graphicx}
\usepackage{dcolumn}
\usepackage{bm}
\usepackage[dvipsnames]{xcolor}
\usepackage{mathptmx}

\usepackage{etoolbox}
\usepackage{braket}
\usepackage{dsfont}

\usepackage[mathlines]{lineno}

\makeatletter
\def\@email#1#2{%
 \endgroup
 \patchcmd{\titleblock@produce}
  {\frontmatter@RRAPformat}
  {\frontmatter@RRAPformat{\produce@RRAP{*#1\href{mailto:#2}{#2}}}\frontmatter@RRAPformat}
  {}{}
}%
\makeatother
\begin{document}


\title{Modeling the dynamics of quantum systems coupled to large dimensional baths using effective energy states}%
 
\author{Loïse Attal}
\email{loise.attal@universite-paris-saclay.fr}
 \affiliation{Université Paris-Saclay, CNRS, Institut des Sciences Moléculaires d'Orsay, 91405 Orsay, France
 }%
\author{Cyril Falvo}%
\affiliation{Université Paris-Saclay, CNRS, Institut des Sciences Moléculaires d'Orsay, 91405 Orsay, France
}%
\affiliation{%
	Université Grenoble-Alpes, CNRS, LIPhy, 38000 Grenoble, France
}%

\author{Florent Calvo}
\affiliation{Université Grenoble-Alpes, CNRS, LIPhy, 38000 Grenoble, France
}

\author{Pascal Parneix}
\email{pascal.parneix@universite-paris-saclay.fr}
\affiliation{Université Paris-Saclay, CNRS, Institut des Sciences Moléculaires d'Orsay, 91405 Orsay, France
}%

\date{\today}

\begin{abstract}
The quantum dynamics of a low-dimensional system in contact with a large but finite harmonic bath is theoretically investigated by coarse-graining the bath into a reduced set of effective energy states. In this model, the couplings between the system and the bath are obtained from the statistical average over the discrete, degenerate effective states. Our model is aimed at intermediate bath sizes in which non-Markovian processes and energy transfer between the bath and the main system are important.
The method is applied to a model system of a Morse oscillator coupled to 40 harmonic modes. The results are found to be in excellent agreement with the direct quantum dynamics simulations of Bouakline \textit{et al.} [J. Phys. Chem. A  \textbf{116}, 11118–11127 (2012)], but at a much lower computational cost. Extension to larger baths is discussed in comparison to the time-convolutionless method. We also extend this study to the case of a microcanonical bath with finite initial internal energies. The computational efficiency and convergence properties of the effective bath states model with respect to relevant parameters are also discussed.
\end{abstract}

\maketitle
\section{Introduction}
\label{sec:Intro}

Quantum processes such as tunneling, delocalization, state superposition, or zero-point energy are ubiquitous in physics and chemistry, and are particularly important for systems ranging from atoms and small molecules in gaseous phase to large molecular aggregates and the condensed phases.\cite{Chergui1996,Martin2004} 
The theoretical description of these processes may be formally achieved through a direct resolution of the time-dependent or the time-independent Schrödinger equation using a full dimensional wavefunction that takes into account all degrees of freedom.\cite{QD:2007uq,bowman_variational_2008,Gatti2009}
Such an approach is in principle exact but only feasible for relatively small systems as it suffers from the so-called curse of dimensionality which arises from the exponential increase in the dimension of the Hilbert space with increasing system size. Despite such difficulties, many methods have been developed to accurately treat systems in a wavefunction approach, including vibrational self-consistent field (VSCF),\cite{Bowman_SCF-SI_1979,bowman_variational_2008,Roy_pccp_2013,Qu_review_2019_pccp} vibrational configuration interaction (VCI),\cite{Bowman_SCF-SI_1979,bowman_variational_2008,Hanchao_2014,Qu_review_2019_pccp} vibrational coupled cluster (VCC),\cite{Christiansen:2007fk} as well as the multiconfigurational time-dependent Hartree (MCTDH) method.\cite{MEYER199073,manthe_wavepacket_1992} While being very accurate, these approaches are also computationally expensive. 
The more recent multi-layer version\cite{wang_multilayer_2003,manthe_multilayer_2008} of the MCTDH method does not suffer from the exponential scaling with system size and allows for larger systems to be reached but it is still quite involved, making it unpractical when the initial states require extensive sampling, as is the case at finite temperatures.\cite{Dunnett_2021_fchem} Hence, most wavefunction-based methods are employed at vanishing temperature. \cite{manthe_wavepacket_1992,Nest_Meyer_2003,Morse_Surface_2012,fischer_hierarchical_2020}

However, in many situations it is only necessary to describe a small part of the system explicitly, the remaining degrees of freedom being treated as an environment (often called a bath) that can be described with appropriate approximations. Many system-bath methods have been developed to describe open quantum systems, by including environmental effects such as dissipation, dephasing or decoherence.\cite{CALDEIRA1981,Breuer:2007ab,Jovanovski_Nonadiabatic_2023}
Such methods are often used within a density matrix approach where only the reduced density matrix of the system is propagated after tracing out the bath.\cite{Breuer:2007ab,Finite_bath_2022} The reduced density matrix follows a quantum master equation that includes terms implicitly representing the bath and its interaction with the system.\cite{Breuer:2007ab} Different approaches have been proposed, differing on how the bath is described, starting with Markovian and perturbative approximations,\cite{Scully_quantum_1967,MOLLOW1969464,Tanimura_review_2020_jcp} as in the Redfield equation \cite{REDFIELD1965} for example. These approximations are particularly useful for very large systems, as in condensed phases.
 
Other methods have been developed beyond perturbative and Markovian treatments,\cite{Dunnett_2021_fchem,Finite_bath_2022,Teixeira_weak-coupling_2022,Vega_non-Markovian_2017} including the time-convolutionless (TCL) approach,\cite{Breuer:2001aa} the auxiliary density matrix approach~\cite{Meier:1999vn} and the  hierarchical equations of motion (HEOM).\cite{Tanimura_review_2020_jcp}
The density matrix renormalization group (DMRG) and its affiliated approaches can also be used to describe the dynamics of open quantum systems.\cite{tamascelli_efficient_2019, Baiardi_DMRG_2020,Prior_Efficient_2010,Dunnett_2021_fchem}   
A complementary approach to reduced density matrix propagation consists in solving the Schrödinger equation by explicitly including the system and the bath but with major simplifications of the bath representation together with some possible dimension reduction.
While inheriting the dimensional issues from the full dimensional wavefunction approach, such methods allow the representation of larger environments by assuming simple bath structures (\textit{e.g.} baths made from uncoupled harmonic oscillators\cite{fischer_hierarchical_2020} or two-level systems\cite{Baer_Quantum_1997}), reducing the number of degrees of freedom using effective modes\cite{fischer_hierarchical_2020} or introducing semiclassical approximations.\cite{Mukamel_Femtosecond_1990,Fried_chapter_1993}
Although such techniques can only represent finite dimensional baths, they aim at reproducing the purely dissipative and irreversible behavior of an infinite bath, the challenge being to converge toward such behavior with the fewest possible degrees of freedom. 

System-bath methods in general have been applied to a large variety of phenomena. Historically starting from the line shape analysis of nuclear magnetic resonance,\cite{Wangsness_dynamical_1953} maser and laser spectra,\cite{Scully_quantum_1967} they are still actively used to investigate the influence of dissipation on coherent laser control \cite{aerts_laser_2022,ramos_ramos_manipulating_2022} or non-Markovian effects in ultrafast nonlinear spectroscopy.\cite{Tanimura:2006fk,Tanimura_review_2020_jcp,Falvo:2015ty}
More generally, such methods naturally apply whenever a small system is in contact with a larger, possibly infinite environment, as in the adsorption\cite{Morse_Surface_2012,bouakline_quantum-mechanical_2019} or chemisorption\cite{Bonfanti_2015_jcpI,Bonfanti_2015_jcpII} of atoms or molecules on surfaces, in electron-phonon couplings in solids,\cite{Mason_QMC_1989} 
or for charge transfer in the condensed phase.\cite{Yan_Solvation_1988,wang_semiclassical_1999,Craig_2007_jcp} 
 
All the aforementioned methods were conceived in the context of an infinite bath, and the case of intermediate-sized baths, too large to be treated exactly through a full dimensional wavefunction approach but too small to be considered as unperturbed by the main system, remains essentially overlooked. However, it is important for a broad variety of problems, ranging from internal energy redistribution in polyatomic molecules,\cite{Gelbart_Random_2003} aggregation and fragmentation processes in finite clusters,\cite{Gross_book_2000,Junghans_Microcanonical_2006} to the effects of nano or microscale environments in the context of quantum computing.\cite{Benenti_Emergence_2001,halbertal_nanoscale_2016,Pekola_finite-size_2016}
In this context, Esposito and Gaspard\cite{esposito_quantum_2003,esposito_spin_2003,esposito_quantum_2007} introduced a microcanonical master equation that takes into account the influence of the system's evolution on the state of a finite bath, together with a rigorous conservation of the total system-bath energy. A similar approach was recently developed and used to describe the nonequilibrium dynamics of the central spin model.\cite{Riera-Campeny:2021aa,Finite_bath_2022} 

In the present contribution, an alternative model is developed based on a system-bath approach, in which the finite bath of harmonic oscillators is replaced by a single ladder of effective energy states representing the total energy stored inside the bath. In this coarse-grained treatment of the bath, which will be denoted as the effective bath states (EBS) approach, all couplings between the system and individual bath modes are fully taken into account.
Within the EBS model, it is possible to keep some information about the global state of the bath itself and to follow its time evolution. The main advantage of transforming the multiple bath modes into a single ladder of effective energy states is to strongly reduce the dimensionality of the problem under scrutiny and its scaling with the number of bath degrees of freedom, allowing for large systems to be considered. The use of effective energy states also enables the bath to be directly prepared at a given, possibly nonzero energy, thus facilitating the simulation of finite temperature effects in relatively large baths by avoiding an expensive sampling of the initial states. 

The derivation of the EBS model is detailed in Sec. \ref{sec:Theory}.
Section \ref{sec:Application} is dedicated to the application of the method to a model system taken from Bouakline and coworkers,  \cite{Morse_Surface_2012} which consists of a 1D Morse potential coupled to 40 harmonic oscillators mimicking a physical surface. Our results are found to agree quantitatively with the MCTDH calculations of these authors. The effects of increasing the initial energy of the bath or its size on the relaxation dynamics of the first vibrational excited state of the Morse oscillator are also quantitatively addressed, notably in comparison with the predictions of the TCL approach.
Finally, some conclusions and perspectives are given in Sec. \ref{sec:conclusion}.

\section{Theory}
\label{sec:Theory}

We consider a one-dimensional system with coordinate $\hat{z}$ and associated momentum $\hat{p}_z$, interacting with a $g$-dimensional bath described by its own positions $\hat{\bf{x}} = \{\hat{x}_1,\hdots, \hat{x}_g\}$ and momenta $\hat{\bf{p}} = \{\hat{p}_1,\hdots, \hat{p}_g\}$.

\subsection{System-bath Hamiltonian}

Following the standard system-bath conventions, the Hamiltonian $\hat{H}$ is divided into three parts associated with the system ($\hat{H}_{\rm{S}}$), the bath ($\hat{H}_{\rm{B}}$), and the system-bath interaction ($\hat{H}_{\rm{SB}}$), respectively:
\begin{equation}
\hat{H}(\hat{z},\hat{p}_z,\hat{\bf{x}},\hat{\bf{p}}) = \hat{H}_{\rm{S}}(\hat{z},\hat{p}_z) + \hat{H}_{\rm{B}}(\hat{\bf{x}},\hat{\bf{p}}) +\hat{H}_{\rm{SB}}(\hat{z},\hat{\bf{x}}).
\label{eq:H_tot}
\end{equation}
Without loss of generality, the system Hamiltonian is given by the sum of its kinetic and potential energy operators
\begin{equation}
\hat{H}_{\rm{S}}(\hat{z},\hat{p}_z) =\frac{\hat{p}_z^2 }{2\mu}+ \hat{V}(\hat{z}),
\label{eq:Hsystem}
\end{equation} 
with $\hat{p}_z = -i\hbar \frac{\partial}{\partial z}$ and $\mu$ the reduced mass of the system.
The diagonalization of this Hamiltonian gives the eigenstates $\ket{v}$ and the corresponding eigenenergies $E_v$. 

Inspired by normal mode analysis, the bath modes are treated as a set of $g$ uncoupled harmonic oscillators of frequencies $\{\omega_k,\; k = 1, \hdots ,g\}$, resulting in the following bath Hamiltonian
\begin{equation}
\hat{H}_{\rm{B}}(\hat{\bf{x}},\hat{\bf{p}}) =\sum_{k = 1}^g  \frac{\hat{p}_k^2}{2\mu_k}+ \frac{\mu_k\omega_k^2}{2}\hat{x}_k^2,
\label{eq:Hbath}
\end{equation}
with $\hat{p}_k = -i\hbar \frac{\partial}{\partial x_k}$ and $\mu_k$ the reduced mass of mode $k$.
The eigenstates of an individual harmonic bath mode $k$ will be denoted as $\ket{n_k}$. Since the bath is made of uncoupled oscillators, it can be naturally described using the uncoupled states $\ket{\bf n}~=~ \ket{n_1}\otimes \ket{n_2}\otimes ...\otimes \ket{n_g}$, which will be referred to as the microstates of the bath. 

Finally, the system-bath coupling is given by the following Hamiltonian
\begin{equation}
\hat{H}_{\rm{SB}}(\hat{z},\hat{\bf{x}}) = f(\hat{z})\times \sum_{k = 1}^g c_k \hat{x}_k.
\label{eq:Hcoupling}
\end{equation}
Since Caldeira and Leggett's seminal work,\cite{CALDEIRA1981,CALDEIRA1983} microscopic models for dissipation usually consider a system-bath coupling that is linear in both the system and bath coordinates. This corresponds to a weak coupling approximation in the sense that the system and bath respond linearly to the perturbation of each other.\cite{CALDEIRA1981}
However, when treating the possibly large amplitude motion of the system, as in the formation or breaking of chemical bonds, adsorption and desorption at a surface, or for large amplitude vibrational modes, the linearity of the coupling in the system coordinate becomes questionable.\cite{NEST200165}
Therefore we lift this constraint and consider an arbitrary function $f(\hat{z})$ of the system coordinate.

\subsection{Effective energy states for the bath}
\label{Bath}

\begin{figure*}[t]
	\includegraphics[width=16cm]{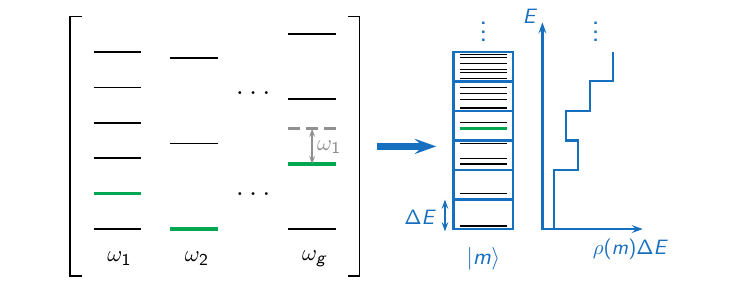}
	\caption{\label{fig:model2}Pictorial representation of the coarse-graining procedure of the bath. Left: energy levels of the harmonic bath modes with frequencies $\omega_1,\omega_2,\hdots,\omega_g$.  Right: Transformation of these modes into a single ladder of effective energy states $\ket{m}$ (in blue) characterized by their energy $m\Delta E$ and density of states $\rho(m)$. Individual microstates are represented in black inside the effective state that contains them. One specific bath state with only two energy quanta in modes 1 and $g$ is highlighted in green (left). The position of the associated microstate inside the effective bath states ladder, obtained by summing the energies of the individual bath modes, is also represented in green (right).}
\end{figure*}

If treated as an ensemble of $g$ uncoupled quantum harmonic oscillators that have access to $N$ energy levels each, the description of the bath still requires typically $N^g$ states. To avoid such exponential scaling, the quantum bath states are transformed into a single ladder of effective states describing the total bath energy.
A given microstate of the bath is defined by a set of quantum numbers ${\bf n}~=~\{n_1, n_2, ..., n_g\}$, which give the number of quanta of energy in each bath mode. The vibrational energy associated to such a state is written as
\begin{equation}
 E({\bf n}) = \sum_{i = 1}^g n_i\hbar \omega_i, 
\end{equation}
and the bath Hamiltonian is given by

\begin{equation}
\hat{H}_{\rm{B}}
=\sum_{{\bf n}} E({\bf n}) \ket{{\bf n}}\bra{{\bf n}}.
\label{eq:bath_exact}
\end{equation}
However, instead of using the complete set of microstates that are too numerous to be handled numerically, the bath is described by a finite number $M$ of effective states $\ket{m}$ that quantize the total energy inside the bath. 
By discretizing the bath energy using an energy grain $\Delta E$, a given effective state $\ket{m}$ contains by definition all microstates $\ket{\bf n}$ such that
\begin{equation}
m\Delta E 	\leq E({\bf n}) < (m+1)\Delta E.
\label{eq:def_bin}
\end{equation}
In the following we write ${\bf n}\in m$ to denote that $\ket{\bf n}$ satisfies Eq.~(\ref{eq:def_bin}). Such states will be considered to have an energy $m\Delta E$. Denoting $\rho$ the density of states (DOS) of the bath, an effective state $\ket{m}$ contains $\rho(m) \times \Delta E$ microstates $\ket{\bf n}$. 
The transformation from a bath of $g$ harmonic modes to a coarse-grained ladder of effective energy states with energy $m\Delta E$ and DOS $\rho(m)$  is displayed in Fig.~\ref{fig:model2}.
The coarse-graining procedure implicitly assumes that all effective states within any given energy bin are equiprobable. This assumption itself implies that energy redistribution within the bath variables is extremely fast compared to the exchange rate constants between the system and the bath, thereby ensuring the validity of microcanonical statistics at all times. The effective energy states can thus be considered to fundamentally carry some stochastic character. 

To derive an effective Hamiltonian using the energy bath states $\ket{m}$, the sum of Eq.~(\ref{eq:bath_exact}) is reordered by increasing energy. Since summing over all possible states ${\bf n}$ is equivalent to summing over all states in a given energy bin $m$ and then summing over all bins $m$, and given that all states ${\bf n}\in m$ are considered to have an energy $m\Delta E$, the bath Hamiltonian of Eq.~(\ref{eq:bath_exact}) can be written as

\begin{align}
\hat{H}_{\rm{B}}
&= \sum_m \sum_{{\bf n}\in m} E({\bf n})  \ket{{\bf n}}\bra{{\bf n}}, \nonumber \\
&= \sum_m m\Delta E\sum_{{\bf n}\in m}  \ket{{\bf n}}\bra{{\bf n}}.
\label{Hb_construction}
\end{align}
Due to the assumed equiprobability of all microstates ${\bf n}\in m$, each effective energy state $|m\rangle$ appears
as degenerate and contains $\rho(m)\Delta E$ equivalent microstates. 
A state $\ket{{\bf n}}$ in the Hamiltonian above is thus considered as a representation of the effective state $\ket{m}$ chosen with a probability $\frac{1}{\rho(m)  \Delta E}$ and hence replaced by
$ \frac{1}{\sqrt{\rho(m)  \Delta E}} \ket{m}.$
This leads to rewrite the projector $\sum_{{\bf n}\in m}  \ket{{\bf n}}\bra{{\bf n}}$ as 
\begin{equation}
\sum_{{\bf n}\in m} \frac{1}{{\rho(m) \Delta E}} \ket{m}\bra{m} = \ket{m}\bra{m}.
\end{equation}

The system-bath basis set $\ket{v,m} = \ket{v}\otimes\ket{m}$ is defined using the system eigenstates $\ket{v}$ and the effective bath states $\ket{m}$. Using this basis set and the closure relations $\sum_m \ket{m}\bra{m}= \sum_{\bf n}\ket{\bf n}\bra{\bf n}=\mathds{1}$ and $\sum_v \ket{v}\bra{v}=\mathds{1}$, the effective Hamiltonians for the system and the bath, $\hat{H}_{\rm S}$ and $\hat{H}_{\rm B}$, can be respectively written as
\begin{align}
&\hat{H}_{\rm{S}} =  \sum_{v= 0}^{N_v-1}\sum_{m = 0}^{M-1} E_v \, \ket{v,m}\bra{v,m},
\label{eq:Hs_model}\\
&\hat{H}_{\rm{B}} =  \sum_{v= 0}^{N_v-1}\sum_{m = 0}^{M-1} m\Delta E \, \ket{v,m}\bra{v,m}.
\label{eq:Hb_model}
\end{align}
The system-bath coupling necessarily induces transitions inside the bath. Therefore, and as will be shown in the next section, the coupling itself can be coarse-grained using the effective energy states, allowing to formally write it using the system-bath basis set $\ket{v,m}$ as 
\begin{equation}
\hat{H}_{\rm{SB}} =  \sum_{v,v'}\sum_{m,m'}\bra{v',m'}f(\hat{z})\sum_{k}c_k\hat{x}_k\ket{v,m}\ket{v',m'}\bra{v,m},
\label{eq:Hsb}
\end{equation}
and allowing the original basis set $\ket{v}\otimes\ket{n_1}\otimes \ket{n_2}\otimes \hdots\otimes \ket{n_g}$ of size $\sim N_v\times N^g$ to be traded for an effective basis set of size $N_v \times M$, with $M\ll N^g$.


\subsection{Coupling Hamiltonian in the system-bath basis set}
\label{subsec:H_coupling}

Having defined the effective bath states, the next objective is to write the coupling Hamiltonian $\hat{H}_{\rm{SB}}$ in the associated system-bath basis set. The main challenge here lies in that the effective states $\ket{m}$ are global states delocalized over the whole bath, whereas the coupling terms of Eq.~(\ref{eq:Hcoupling}) only involve one bath mode.
The way by which the interaction between the system and a given bath mode is accounted for will be detailed by focusing on a specific mode $k$ in $\hat{H}_{\rm{SB}}$ and by computing the  matrix element
\begin{equation}
\bra{v',m'}f(\hat{z})c_k\hat{x}_k\ket{v,m} = \bra{v'}f(\hat{z})\ket{v}\times c_k  \bra{m'}\hat{x}_k\ket{m}.
\end{equation}
The calculation of $\bra{v'}f(\hat{z})\ket{v}$ is straightforward and it mostly remains to express the bath part of this coupling term, \textit{i.e.} the matrix elements of operator $\hat{x}_k$ in the effective bath states $\ket{m}$.

\subsubsection{Partition of the bath energy}

Assuming that the bath is at a given energy $m\Delta E$ and isolating mode $k$ from the other bath modes, the energy can be expressed as
\begin{equation}
m\Delta E = n_k \hbar \omega_k + E^*_{n_k},
\label{eq:bin_m}
\end{equation}
where $n_k \hbar \omega_k$ is the (harmonic) energy of mode $k$ when there are $n_k$ quanta of energy in this mode, and $E^*_{n_k}$ is the energy accessible by the other bath modes, which is given by 

\begin{equation}  
E^*_{n_k} = m\Delta E-n_k \hbar \omega_k = \sum_{j \ne k} n_j\hbar \omega_j.
\label{eq:Spectator_nrj}
\end{equation}
For a given value of $m$ and $n_k$ there are multiple possibilities to distribute $E^*_{n_k}$ between the remaining modes of the bath.
These remaining modes will be called spectator modes as they are not modified by operator $\hat{x}_k$.
A bin $m^*$ can be associated to $E^*_{n_k}$ and the expression ${\bf n}^* \in m^*$ will be used to indicate that the (spectator) microstate ${\bf n}^* = \{ n_j\}_{j\ne k}$ satisfies Eq.~(\ref{eq:Spectator_nrj}).
 
Operator $\hat{x}_k$ only affects mode $k$ for which it induces a transition between $n_k$ and $n'_k~=~n_k~\pm~1$. Hence, the energy $E^*_{n_k}$ contained in the spectator modes $j\ne k$ is not affected by the transition and the bath energy becomes
\begin{equation}
m'\Delta E = n'_k \hbar \omega_k + E^*_{n_k}.
\label{eq:bin_m_prime}
\end{equation}
From Eqs.~(\ref{eq:bin_m}) and (\ref{eq:bin_m_prime}) it follows that

\begin{equation}
m\Delta E-n_k \hbar \omega_k = E^*_{n_k}= m'\Delta E-n'_k \hbar \omega_k.
\label{eq:E_star}
\end{equation}
Therefore the transitions induced by $\hat{x}_k$ are characterized by a unique $\Delta m_k~=~|m' - m|$ that does not depend on $n_k$ and $E^*_{n_k}$, but only on $\omega_k$ and $\Delta n_k~=~ n'_k - n_k$, which is equal to $\pm 1$ for such a linear coupling. Hence, a state $\ket{m}$ is coupled to the states $\ket{m+\Delta m_k}$ and $\ket{m-\Delta m_k}$ in such a way that 
\begin{equation}
\Delta m_k = \left|\frac{\Delta n_k \hbar \omega_k}{\Delta E}\right|= \frac{\hbar \omega_k}{\Delta E}.
\label{eq:def_delta_m}
\end{equation}
As seen in the expression above, $\Delta m_k$ depends on the chosen mode $k$ through its frequency $\omega_k$, however for the sake of simplifying the notations we will omit the subscript $k$ in the remainder of the discussion.

\begin{figure*}
	\includegraphics[width=16cm]{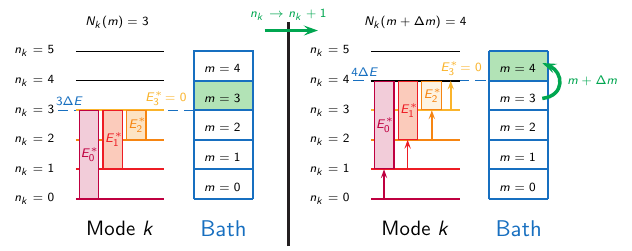}
	\caption{\label{fig:transitions} Pictorial representation of the method used to determine the bath part of the coupling terms. As an example, the coupling between $m = 3$ (left side) and $m +\Delta m = 4$ (right side) due to transitions $n_k \rightarrow n_k +1$ of a given mode $k$ with $\hbar\omega_k = \Delta E$ is represented. For each harmonic level $n_k$ accessible in $m$ the remaining energy $E^*_{n_k}$ is represented by a box of the same color as the said level. This energy is shared between the spectator bath modes, which have been omitted for clarity. On the right side, colored arrows represent transitions  $n_k \rightarrow n_k +1$. The maximum values of $n_k$ accessible in both cases are also indicated.}
\end{figure*}

Only one of the possible values of $n_k$ was considered so far. To obtain the total coupling between $\ket{m}$ and $\ket{m\pm\Delta m}$ the contributions of all accessible states $n_k$ in $\ket{m}$ must be added, meaning all the $n_k$ such that  $n_k\hbar \omega_k < (m+1)\Delta E$. Defining $N_k(m)$ as the maximum integer fulfilling this condition for a given $m$, the total coupling is obtained by considering all $n_k$ such that $0\leq n_k \leq N_k(m)$. 
This procedure is shown in Fig.~\ref{fig:transitions} using a case where $\hbar\omega_k = \Delta E$ was chosen for clarity reasons, the bath being assumed to lie in $m=3$.
The amount of energy $E^*_{n_k}$ in the spectator modes is not modified by the $n_k \rightarrow n_k +1$ transition and $E^*_{n_k}$ is only shifted on the energy scale when $m\Delta E$ becomes $(m+\Delta m)\Delta E$. Since $\hbar\omega_k = \Delta E$, the transitions from a given $n_k$ to $n_k+1$ are all characterized by  $\Delta m = 1$ and the individual transitions depicted in Fig.~\ref{fig:transitions} all end at the same energy $(m+\Delta m)\Delta E=4\,\Delta E$.
The figure also illustrates the definition of $N_k(m)$ and shows that for the $n_k \rightarrow n_k +1$ transition, the final bin is characterized by $N_k(m+\Delta m) = N_k(m) +1$.

\subsubsection{Rounding the frequencies}
\label{Rounded frequencies}

Coarse-graining the bath into a reduced number of effective energy states entails a careful examination of how these states should be identified upon arbitrary transitions. In particular, the choice of $\hbar\omega_k$ as an exact multiple of $\Delta E$ in the above example is not as restrictive as it may appear at first sight, and rounding the bath frequencies to the nearest multiple of $\Delta E$ yields a correct procedure. To show this, we note that
Eqs.~(\ref{eq:bin_m}), (\ref{eq:bin_m_prime}) and (\ref{eq:def_delta_m}) all
implicitly assume that $E({\bf n}) =n_k \hbar \omega_k + E^*_{n_k}$ is a multiple of $\Delta E$. 
Satisfying these equations could be ensured by rounding the bath energies to the nearest multiple of $\Delta E$, but it is important that transition energies be correctly coarse-grained too. In particular, rounding down to the lowest multiple of $\Delta E$ would cause ambiguities in the assignment of the final bin of a transition. This can be illustrated by considering the integer $m$ obtained by rounding the bath energy down to the lowest multiple of $\Delta E$,
\begin{equation}
m= {\rm Int}\!\left(\frac{n_k \hbar \omega_k + E^*_{n_k}}{\Delta E} \right),
\end{equation}
where Int denotes the lowest immediate integer. For a transition $m\to m'=m\pm\Delta m$, the corresponding coarse-grained transition integer $\Delta m$ would also need to be rounded down and be written as
\begin{equation}
\Delta m = {\rm Int}\!\left( \frac{\hbar \omega_k}{\Delta E}\right).
\end{equation}
However, the final bin $m'$ of the transition would then become ambiguous, with two nonequivalent definitions namely as
\begin{align}
m'&= {\rm Int}\!\left(\frac{(n_k \pm1)\hbar \omega_k + E^*_{n_k}}{\Delta E} \right) \nonumber\\
&= {\rm Int}\!\left(\frac{n_k\hbar \omega_k + E^*_{n_k}}{\Delta E} \pm\frac{\hbar \omega_k}{\Delta E}\right), 
\label{eq:m_prime}
\end{align}
but also as 
\begin{equation}
m \pm \Delta m = {\rm Int}\!\left(\frac{n_k\hbar \omega_k + E^*_{n_k}}{\Delta E}\right)\pm{\rm Int}\!\left( \frac{\hbar \omega_k}{\Delta E}\right).
\label{eq:m_plus_dm}
\end{equation}
Since rounding down a sum of two terms is not equivalent to rounding down each term before summing them, the two expressions above differ. This could cause errors in assigning properly the final bin of some transitions, and contribute to some important resonant processes being missed.

Instead, rounding the bath frequencies, up or down, to the nearest multiple of $\Delta E$ ensures that all the quantities involved in the equations above are true integers, and more importantly that
\begin{equation}
\hbar \omega_k = m_k \Delta E
\end{equation}
always exactly holds. If $\Delta E$ is sufficiently small, this approximation does not affect much the values of the frequencies. Using the rounded bath frequencies, the energy of a given microstate of the bath becomes
\begin{equation}
E({\bf n}) = \sum_{i = 1}^g n_i\hbar \omega_i
=\sum_{i = 1}^g n_i m_i \Delta E 
= m  \Delta E, 
\end{equation}
and can be split as in Eq.~(\ref{eq:bin_m}) to obtain
\begin{align}
m \Delta E &= n_k m_k \Delta E + \sum_{j \ne k} n_j m_j \Delta E \nonumber\\
&= n_k m_k \Delta E + m^*\Delta E
\end{align}
with 
\begin{equation}
m^* =\sum_{j \ne k} n_j m_j = m-n_km_k.
\label{m_star}
\end{equation}
Similarly, the transition is defined by $\Delta m = m_k$ and the final bin by $(m\pm \Delta m) \Delta E = (n_k\pm1) m_k \Delta E+ m^*\Delta E$. 
The rounding procedure also gives a simple way to compute the maximum number of accessible quanta for mode $k$ inside a bin $m$, which is now the integer $N_k(m)$ such that
\begin{equation}
N_k(m)\times m_k = m.
\end{equation}
Hence $N_k(m)$ is reached when there is no energy left for the spectator modes, as is the case in Fig.~\ref{fig:transitions}.

\subsubsection{Coupling terms between effective states}

To compute the coupling Hamiltonian in the system-bath basis set $\ket{v,m}$, the operator $\hat{x}_k$ is first written in the exact basis set of the bath as
\begin{align}
\hat{x}_k &= \sum_{\bf n}\sum_{\bf n'}\bra{\bf n'} \hat{x}_k \ket{\bf n} \ket{{\bf n'}}\bra{{\bf n}}\nonumber\\
 &= \sum_{m}\,\sum_{m'}\,\sum_{{\bf n}\in m}\,\sum_{{\bf n'}\in m'}\bra{{\bf n'}} \hat{x}_k \ket{{\bf n}} \ket{{\bf n'}}\bra{{\bf n}}.
\end{align}
Writing $\ket{\bf n}$ as $\ket{n_k, {\bf n}_{j\ne k}}= \ket{n_k,{\bf n}^*}$ to isolate mode $k$ from the other bath modes, the action of $\hat{x}_k$ on the states is given by
\begin{align}
\bra{{\bf n'}} \hat{x}_k \ket{{\bf n}} &= \bra{n'_k,{\bf n'}^*} \hat{x}_k \ket{n_k,{\bf n}^*} \nonumber \\
&=  \bra{n'_k} \hat{x}_k \ket{n_k}\braket{{\bf n'}^* \mid {\bf n}^*}  \nonumber\\
&= \bra{n'_k} \hat{x}_k \ket{n_k}\times\delta_{n'_k, n_k \pm 1}\times\delta_{{\bf n'}^*,{\bf n}^*}.
\end{align}
Using the equation above and the following reordering of the sums
\begin{equation}
\sum_{{\bf n}} \ket{\bf n} = \sum_{m = 0}^{M-1} \sum_{{\bf n}\in m} \ket{\bf n}= \sum_{m = 0}^{M-1} \sum_{n_k = 0}^{N_k(m)} \sum_{{\bf n}^* \in m^*}\ket{n_k,{\bf n}^*},
\end{equation}
operator $\hat{x}_k $ can be rewritten as
\begin{align}
 \hat{x}_k 
 =& \sum_{m=0}^{M'-1}\sum_{n_k = 0}^{N_k(m)} \sum_{{\bf n}^*\in m^*}\bra{n_k+1} \hat{x}_k \ket{n_k} \ket{n_k+1,{\bf n}^*}\bra{n_k,{\bf n}^*} \nonumber\\
 &+\sum_{m=\Delta m}^{M-1} \sum_{n_k = 1}^{N_k(m)}  \sum_{{\bf n}^*\in m^*}\bra{n_k-1} \hat{x}_k \ket{n_k} \ket{n_k-1,{\bf n}^*}\bra{n_k,{\bf n}^*} \nonumber\\
 =& \;\hat{x}_{k}^{+} +\hat{x}_{k}^{-},
 \label{eq:x_k}
\end{align}
where $M' = M - \Delta m$. Operators $\hat{x}_{k}^{\pm}$ are associated to transitions $n_k \rightarrow n_k\pm 1$. They will be treated separately, starting with $\hat{x}_{k}^{+}$. 
This coupling operator involves microstates $\ket{n_k,{\bf n}^*} \in m$ and $\ket{n_k +1,{\bf n}^*} \in m'$.
As in Sec.~\ref{Bath}, the state $\ket{n_k,{\bf n}^*}$ is seen as a representation of the effective state $\ket{m}$, chosen among its $\rho(m)\Delta E$ microstates, and thus replaced by
 $\frac{1}{\sqrt{\rho(m)  \Delta E}} \ket{m}$.
For state $\ket{n_k +1,{\bf n}^*}$ it follows from 
\begin{equation}
(n_k+1) m_k \Delta E+ m^*\Delta E =  (m + \Delta m) \Delta E
\end{equation}
that $m' = m + m_k =  m+\Delta m $.
However, only a subset of the microstates in $\ket{m +\Delta m}$ may be written as $\ket{n'_k = n_k +1,{\bf n}^*}$. Notably, a microstate of this form cannot describe a state where $n'_k = 0$. 
Once a microstate $\ket{n_k,{\bf n}^*}$ is randomly chosen inside $\ket{m}$, the transition $n_k \rightarrow n'_k = n_k + 1$ is determined unequivocally. Hence, only a subset of $\rho(m)\Delta E$ microstates inside $\ket{m +\Delta m}$ may be reached through this transition, and state $\ket{n_k +1,{\bf n}^*}$ is replaced by $\frac{1}{\sqrt{\rho(m)  \Delta E}} \ket{m +\Delta m}$.

As an example, once bin $m=3$ is chosen in the left panel of Fig.~\ref{fig:transitions} it sets the number of accessible microstates, which is determined by $N_k(m)$ and by the size of the rectangles representing spectator modes energy $E^*_{n_k}$. The transition $n_k \rightarrow n_k+1$ shifts those quantities along the energy scale but does not change the number of microstates involved, since (i) the spectator modes energy is not modified; and (ii) the number of truly accessible values of $n_k$ remains the same as $n_k$  varies between 0 and $N_k(m) = 3$ in the left panel, and between 1 and $N_k(m+\Delta m) = 4$ in the right panel. Hence, the DOS of bin $m = 3$ entirely determines the transition.

This leads to replacing $\ket{n_k+1,{\bf n}^*}\bra{n_k,{\bf n}^*}$ by $\frac{1}{{\rho(m)\Delta E}} \ket{m+\Delta m}\bra{m}$ in the expression of $\hat{x}_{k}^{+}$.
Introducing the DOS $\rho^{(k)}(m^*)$ such that $\rho^{(k)}(m^*)\Delta E$ is the number of ways to obtain the bath state $\ket{m^*}$ using only spectator bath modes $j\neq k$, this leads to 
\begin{align}
 \sum_{{\bf n}^*\in m^*} \ket{n_k+1,{\bf n}^*}\bra{n_k,{\bf n}^*} \rightarrow \frac{\rho^{(k)} (m^*)}{\rho(m)}\ket{m +\Delta m}\bra{m},
\end{align}
and the effective representation of $\hat{x}_{k}^{+}$ becomes
\begin{equation}
\hat{x}_{k}^{+} = \sum_{m = 0}^{M'-1}  \sum_{n_k = 0}^{N_k(m)}\bra{n_k+1} \hat{x}_k \ket{n_k}  \frac{\rho^{(k)} (m^*) }{\rho(m)} \ket{m+\Delta m}\bra{m}.
\end{equation}
In the expression above, the ratio of the two DOSs corresponds to the probability for a microstate inside $\ket{m}$ to have $n_k$ quanta of energy in mode $k$. Since $m^*$ can be expressed using only $m$ and $n_k$ [see Eq.~(\ref{m_star})], this probability will be denoted as $\mathbb{P}(m,n_k)$. 
Given that $\bra{n_k+1} \hat{x}_k \ket{n_k}=\sqrt{{\hbar (n_k+1)}/{2\omega_k}}$, an explicit expression of $\hat{x}_{k}^{+}$ is obtained as
\begin{equation}
\hat{x}_{k}^{+}  = \sum_{m = 0}^{M'-1}  \sum_{n_k = 0}^{N_k(m)} \sqrt{\frac{\hbar (n_k+1)}{2\omega_k}} \; \mathbb{P}(m,n_k) \ket{m+\Delta m}\bra{m}.
\label{eq:transition_plus}
\end{equation}
Note that the sum over $m$ stops at $M'-1 = M - 1-\Delta m$ as $m+\Delta m$ must not exceed  $M-1$, the highest bin of the bath.

Since $\hat{x}_{k}^{-}$ and $\hat{x}_{k}^{+}$ are conjugate to each other, the calculation of $\hat{x}_{k}^{-}$ is similar, the main difference being that transitions $n_k \rightarrow n_k-1$ exist only if $n_k\geq1$. Therefore the sums involved in this operator start at $n_k = 1 $ and $m = \Delta m = \hbar \omega_k/\Delta E$ as there must be at least one quantum of energy in mode $k$. 
Hence operator $\hat{x}_{k}^{-}$ is given by
\begin{equation}
 \hat{x}_{k}^{-}= \sum_{m=\Delta m}^{M-1}\sum_{n_k = 1}^{N_k(m)}\bra{n_k-1} \hat{x}_k \ket{n_k} \sum_{{\bf n}^*\in m^*} \ket{n_k-1,{\bf n}^*}\bra{n_k,{\bf n}^*}
 \label{eq:x_k_moins}
\end{equation}
where $\ket{n_k,{\bf n}^*} \in \ket{m}$ and $\ket{n_k-1,{\bf n}^*} \in \ket{m-\Delta m}$. As before, these microstates can be seen as a representation of the effective state  containing them, chosen among the possible microstates of their form.
According to Eq.~(\ref{eq:x_k_moins}), $\ket{n_k,{\bf n}^*}$ does not span all the possible states of $\ket{m}$ since the value of $n_k$ is constrained by $1\leq n_k\leq N_k(m)$. However, $n_k - 1$ varies between  0 and $N_k(m)-1 = N_k(m -\Delta m)$, hence $\ket{n_k-1,{\bf n}^*}$ spans all possible states in $\ket{m -\Delta m}$ and a probability $\frac{1}{{\rho(m-\Delta m)\Delta E}}$ is associated to each microstate involved in Eq.~(\ref{eq:x_k_moins}), leading to the transformation
\begin{equation}
\sum_{{\bf n}^*\in m^*} \ket{n_k-1,{\bf n}^*}\bra{n_k,{\bf n}^*} 
\rightarrow \frac{\rho^{(k)} (m^*)}{\rho(m-\Delta m)}\ket{m -\Delta m}\bra{m}.
\end{equation}
Given that $m^* = m-n_km_k = m -\Delta m - (n_k-1)m_k$, the ratio of DOSs is interpreted as the probability $\mathbb{P}(m-\Delta m,n_k-1)$ of having a microstate with $n_k-1$ quanta in mode $k$ inside $\ket{m-\Delta m}$.
Expressing $\bra{n_k-1} \hat{x}_k \ket{n_k} $ in the harmonic basis set as $\sqrt{{\hbar n_k}{2\omega_k}}$, the effective representation of $\hat{x}_{k}^{-}$ is obtained as
\begin{align}
\hat{x}_{k}^{-}&=\sum_{m = \Delta m}^{M-1}  \sum_{n_k = 1}^{N_k(m)}\sqrt{\frac{\hbar n_k}{2\omega_k}}\; \mathbb{P}(m-\Delta m,n_k-1) \ket{m -\Delta m}\bra{m} \nonumber\\
&=\sum_{m' = 0}^{M'-1}  \sum_{n'_k = 0}^{N_k(m')}\sqrt{\frac{\hbar(n'_k+1)}{2\omega_k}}\;  \mathbb{P}(m',n'_k) \ket{m'}\bra{m'+\Delta m}
\label{eq:transition_minus}
\end{align}
with $n'_k = n_k -1$ and $m' = m -\Delta m$. Comparison of Eqs.~(\ref{eq:transition_plus}) and (\ref{eq:transition_minus}) shows that the representation of $\hat{x}_{k} = \hat{x}_{k}^{+} +~\hat{x}_{k}^{-}$ in the effective basis set of the bath is a Hermitian operator. 

Having expressed the coupling operator $\hat{x}_{k}$ associated to a given bath mode $k$, the coupling Hamiltonian can finally be written in the system-bath basis set $\ket{v,m}$ as
\begin{align}
\hat{H}_{\rm{SB}} =&  \sum_{v,v'}\sum_{m,m'} \bra{v'}f(\hat{z})\ket{v} \sum_k c_k \bra{m'}\hat{x}_k\ket{m}\ket{v',m'}\bra{v,m} \nonumber \\
=& \sum_{v= 0}^{N_v-1}\sum_{v'= 0}^{N_v-1}\bra{v'}f(\hat{z})\ket{v}\\ 
&\times \sum_{k = 1}^g c_k\sum_{m = 0}^{M'-1}\sum_{n_k = 0}^{N_k(m)}\sqrt{\frac{\hbar(n_k+1) }{2\omega_k}}\,\mathbb{P}(m,n_k) \nonumber\\  
&\times\left(\,\ket{v',m +\Delta m_k}\bra{v,m}+\ket{v',m}\bra{v,m +\Delta m_k}\,\right),\nonumber
\label{eq:Hsb_model}
\end{align}
where the notation $\Delta m_k$ was reintroduced [see Eq.~(\ref{eq:def_delta_m})] since several bath modes with different transition parameters are involved in the Hamiltonian. The effective bath states $\ket{m}$ are constructed in such a way that the coupling Hamiltonian above accounts for the average coupling between the system and all microstates contained in a given energy bin by using microcanonical probabilities.

\subsection{Numerical details and data analysis}

Once the total Hamiltonian $\hat{H}$ is obtained, the time-dependent Schrödinger equation (TDSE) is solved starting from an initial state $\ket{v_0,m_0}$, \textit{i.e.} with the system prepared in a given state $v_0$ and the bath starting at a given energy $m_0\Delta E$. The initial state is decomposed in the eigenbasis $\ket{\phi_\alpha}$ of $\hat{H}$ to obtain the initial wavepacket 
\begin{equation}
\ket{\psi(0)}= \ket{v_0,m_0} = \sum_{\alpha} C_\alpha \ket{\phi_\alpha}.
\label{eq:initial_state}
\end{equation}
The effective Hamiltonian being time-independent, the TDSE can be solved exactly to obtain the wavefunction as a function of the eigenenergies $E_\alpha$ of $\hat{H}$  and the coefficients of the initial state
\begin{equation}
\ket{\psi(t)} = e^{-i\hat{H}t/\hbar}\ket{\psi(0)} = \sum_{\alpha} C_\alpha e^{-iE_\alpha t/\hbar}\ket{\phi_\alpha}.
\label{eq:psi_t}
\end{equation}
In practice, not all the eigenstates $\ket{\phi_\alpha}$ are relevant for a given initial condition and only the states such that $|C_\alpha|^2~>~\epsilon~\times~|C_{{\rm{\max}}}|^2$, with $C_{{\rm{\max}}}$ the largest coefficient in the decomposition of $\ket{\psi(0)}$ and $\epsilon$ a threshold parameter, are considered in the calculation. 
The (normalized) populations of the states $\ket{v}$ and $\ket{m}$ are respectively given by 
\begin{align}
{P}_v(t) &= \sum_m |\braket{v,m|\psi(t)}|^2, \nonumber\\ 
 {P}_m(t) &= \sum_v |\braket{v,m|\psi(t)}|^2.
 \label{eq:probas}
\end{align}
The system and bath mean energies can be obtained in a similar fashion.
In practice, in all of the above, the sums over $m$ should only cover values such that $\rho(m) \ne 0$.  If there is no microstate $\ket{\bf n}$ such that $E({\bf n}) = m \Delta E$, then $\ket{m}$ is empty and $\rho(m) = 0$. This is typically the case for effective states associated to low energies. Such empty states have no physical meaning and are excluded from the basis set.

\section{Application: A Morse potential interacting with a bath of harmonic oscillators}
\label{sec:Application}

The EBS model is now applied to a one-dimensional Morse potential coupled to a finite environment modeled by a set of $g =40$--600 harmonic oscillators. This model system was introduced in Ref.~\onlinecite{Morse_Surface_2012}, where the full dimensional dynamics could be characterized for $g=40$ using the MCTDH method. 

\subsection{Morse-surface Hamiltonian}
\label{subsec:Morse_Hamiltonian}

Following Bouakline \textit{et al.}\cite{Morse_Surface_2012} the vibrational Hamiltonian $\hat{H}_{\rm{S}}$ associated to the Morse potential is defined as
\begin{equation}
\hat{H}_{\rm{S}}(\hat{z},\hat{p}_z) =\frac{\hat{p}_z^2}{2\mu} + D\left( e^{-2\alpha \hat{z}} - 2e^{-\alpha \hat{z}} \right).
\end{equation}
As in Ref.~\onlinecite{Morse_Surface_2012}, parameters are chosen to mimic an O-H stretching mode\cite{WRIGHT198515,Morse_Surface_2012} with $ D = 0.1994$ Hartree, $\alpha = 1.189 \, a_0^{-1}$ and $\mu = 0.9481$ amu. This Hamiltonian gives rise to a set of anharmonic eigenstates $\ket{v}$ associated with the eigenenergies $E_v$, which are the vibrational energy levels of the Morse potential. Diagonalization is achieved using the discrete variable representation (DVR) method\cite{DVR_Light_1985,Light_Discrete-variable_2000} with Hermite polynomials. The convergence of the 22 bound states is tested by comparing the numerical eigenvalues to the well-known Morse energies.\cite{Morse_1929} With this choice of parameters,  the harmonic and fundamental frequencies are found at $\omega_h=3954$ cm$^{-1}$ and $\omega_{0 \rightarrow 1} = 3784.5$~cm$^{-1}$, respectively. Anharmonicities also shift the hot bands to the red rather significantly, {\em e.g.} $\omega_{1 \rightarrow 2} =  3605$~cm$^{-1}$ and $\omega_{2 \rightarrow 3} =  3425.5$~cm$^{-1}$.

The bath is described by a standard harmonic Hamiltonian, like the one given by Eq.~(\ref{eq:Hbath}), where we take $\mu_k = 1$ amu for every mode.\cite{Morse_Surface_2012} 
We mainly focus here on the resonant bath of Ref.~\onlinecite{Morse_Surface_2012}, in which the bath frequencies are close to the main transitions of the Morse oscillator, thereby ensuring significant system-bath couplings. The case of the non-resonant bath, also discussed by Bouakline {\em et al.},\cite{Morse_Surface_2012} is briefly addressed in Appendix \ref{AppB}.
In the resonant case, the bath is characterized by uncoupled harmonic oscillators with frequencies given by
 \begin{equation}
 \omega_k = \omega_0 \,+\, k \times\Delta \omega\quad \forall\; k = 1,\hdots,g 
 \end{equation}
Except in Sec.~\ref{subsec:Bath_size}, we follow Ref.~\onlinecite{Morse_Surface_2012} and use $g = 40$ as well as the resonant bath parameters $\omega_0 = 14$ cm$^{-1}$ and $\Delta \omega = 179.5$ cm$^{-1}$. With such parameters $\omega_{21}$ is resonant with the $v = 0 \rightarrow v = 1$ transition, whereas $\omega_{20}$ is resonant with $v = 1 \rightarrow v = 2$. The frequencies of the resonant  bath range from $\omega_1 = 193.5$ cm$^{-1}$ to $\omega_{40} = 7194$ cm$^{-1}$.

Following again Ref.~\onlinecite{Morse_Surface_2012}, the system-bath Hamiltonian is given by
\begin{equation}
\hat{H}_{\rm{SB}}(\hat{z},\hat{\bf{x}})= f(\hat{z})\sum_{k = 1}^g c_k \hat{x}_k = -\frac{1- e^{-\alpha \hat{z}}}{\alpha}\sum_{k = 1}^g c_k \hat{x}_k.
\label{eq:HSB}
\end{equation}
The coupling has an exponential dependency on the Morse coordinate, mimicking a vanishing interaction [constant $f(\hat{z})$] in the dissociating limit.\cite{Morse_Surface_2012} Using an Ohmic bath model, the coupling constants $c_k$ are related to the bath frequencies through\cite{Ohmic_bath_2001,Ohmic_bath_2010, Morse_Surface_2012}
\begin{equation}
c_k = \omega_k \sqrt{\frac{2\mu_k\mu\gamma\Delta\omega}{\pi}},
\end{equation}
where $\gamma$ is the relaxation rate of the system. As in   Ref.~\onlinecite{Morse_Surface_2012}, we use $\gamma^{-1} = 500$ fs.
Note that the exponential decay of the coupling in the system coordinate $\hat{z}$ allows for transitions with any possible value of $\Delta v$, even though transitions $\Delta v = \pm 1$ are favored due to a stronger coupling between a vibrational state and its direct neighbors. In contrast, the linear coupling in the bath coordinates allows only one bath mode at a time to gain or lose exactly one quantum of vibration.

In practice, harmonic densities of states $\rho$ and $\rho^{(k)}$ are computed using the Beyer-Swinehart algorithm.\cite{Beyer_Swinehart_1973} Owing to the preliminary rounding of the bath frequencies required by the coarse-graining method, this counting method is exact if $\Delta E$ is used as the grain size of the algorithm. In order to speedup the calculation, and since for a given value of $m$ the probability $\mathbb{P}(m,n_k)$ is a rapidly decreasing function of $n_k$, only the terms $\mathbb{P}(m,n_k)$ such that $\mathbb{P}(m,n_k)>\varepsilon_p \times \mathbb{P}(m,n_k=0)$, with $\varepsilon_p$ a threshold, are determined. For the calculations reported below, both $\varepsilon$ and $\varepsilon_p$ were set to $10^{-4}$. 

\subsection{Relaxation from the first vibrational excited state}
\label{subsec:v1_m0}

\begin{figure}
  \includegraphics[width=8.5cm]{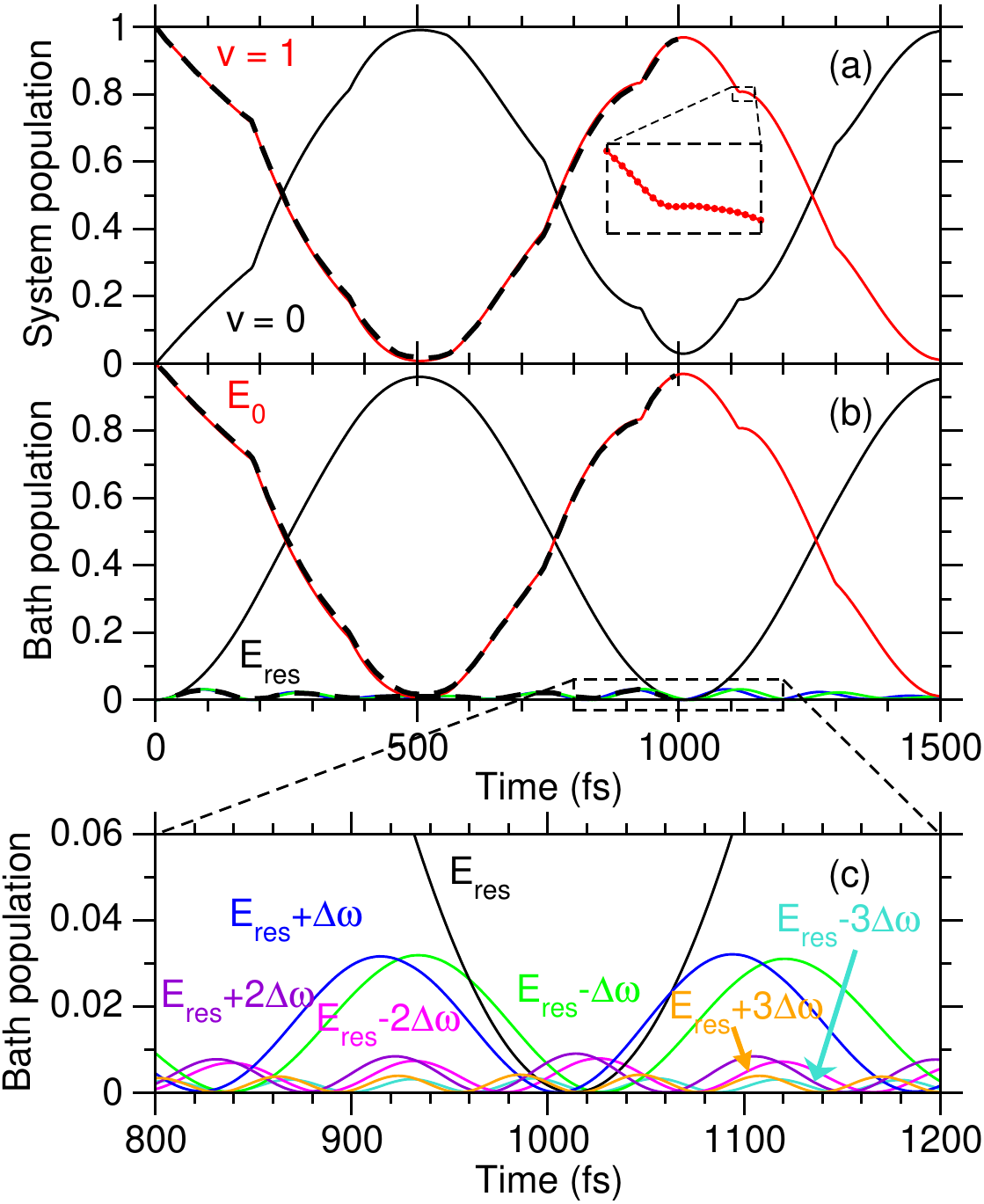}
	\caption{\label{fig:Dyn_v1_m0} Relaxation of a Morse oscillator starting in its first exited state ($v_0 = 1$) and interacting with a resonant bath of 40 harmonic oscillators initially in its ground state ($E_0 = 0$). (a) Time evolution of the population in the Morse states $v$. The inset emphasizes the differentiable character of the populations evolution with time; (b) Time evolution of the population of the effective bath states, labeled according to their energy. The energy $E_{\rm res}$ corresponds to the effective state that is resonant with the $v = 1 \rightarrow v = 0$ transition; (c) Closer view of the bath evolution highlighting the contributions of several off-resonant bath states. In panels (a) and (b), the MCTDH results from Ref.~\onlinecite{Morse_Surface_2012} are superimposed as dashed lines.}
\end{figure}

The reliability of the present EBS model was first tested by following the relaxation of the Morse oscillator when initially prepared in its first vibrational excited state ($v_0 = 1$) and placed in contact with a bath in its ground state ($E_0 = m_0\Delta E = 0$). Starting from $\ket{v_0,m_0} = \ket{1,0}$, the system and the bath populations defined by Eq.~(\ref{eq:probas}) are computed along a 1500 fs trajectory with a temporal resolution of 1~fs.
The results obtained using $N_v = 5$, $M = 6000$ and an energy grain of $\Delta E = 2$ cm$^{-1}$, shown in Fig.~\ref{fig:Dyn_v1_m0}, can be directly compared to those from Ref.~\onlinecite{Morse_Surface_2012}. Note that results from Ref.~\onlinecite{Morse_Surface_2012} are only available along a 1~ps trajectory.

The time evolution of the system displayed in Fig.~\ref{fig:Dyn_v1_m0}(a) starts, as expected, with the first excited state losing its population in favor of the vibrational ground state until its population drops to almost zero near $t = 505$~fs. However, this decay is not exponential as it would be for an infinitely large bath. More remarkably, at longer times the population progressively returns to the excited state $v_0 = 1$. This is due to the finite size of the bath and to the very few bath modes that are significantly involved in the dynamics, causing the system's population to follow a Rabi-like oscillation between $v = 1$ and $v = 0$ with an almost complete recurrence at $T_{\rm rec}= 1012$~fs where 97\% of the population is back in the initial state. 

As can be seen in  Fig.~\ref{fig:Dyn_v1_m0}(b), the bath evolution is mostly a mirror of the system evolution, its main feature being an oscillation between the initial bath state $E_0 = 0$ and the state reached by the bath when the resonant bath mode $k = 21$ ($\omega_{21} \approx \omega_{0 \rightarrow 1} $) gains one quantum of energy from the relaxation of the system from $v = 1$ to $v = 0$. The energy associated to this resonant effective state $\ket{m_{\rm res}}$ is denoted $E_{\rm res}$. The effective state $\ket{m_{\rm res}}$  contains the microstate such that $n_{21} = 1$ and $n_k = 0\; \forall\ k \ne 21$, which is responsible for the resonant transition. This resonant exchange between $\ket{v_0 = 1, m_0 = 0}$ and $\ket{v = 0, m = m_{\rm res}}$ resembles a Rabi oscillation between these two states. More precisely, using an off-resonant Rabi model\cite{Morse_Surface_2012,CohenTannoudjiDiuLaloe} the recurrence time of a bath mode $k$ would be given by
\begin{equation}
T^{(k)} = \frac{2\pi\hbar}{\sqrt{4V_k^2 + (\hbar\Delta\omega_k)^2}},
\label{eq:Rabi_T}
\end{equation}
where $ \Delta\omega_k = |\omega_{0 \rightarrow 1} - \omega_k|$ is the detuning between the system and bath frequencies and $V_k = \bra{1, 0}\hat{H}\ket{0, m_k}$ is the coupling strength between the two involved states. The probability $\mathcal{P}$ of residing in state $\ket{v = 0, m = m_k}$ can then be written as
\begin{equation}
\mathcal{P}_{0,m_k}(t) = \frac{4V_k^2}{4V_k^2 + (\hbar\Delta\omega_k)^2}\sin^2\left(\frac{\pi}{T^{(k)}}t\right).
\label{eq:Rabi_P}
\end{equation}
In the present context, the Rabi period for mode $k = 21$ corresponds to the recurrence time identified in Fig.~\ref{fig:Dyn_v1_m0}, and Eq.~(\ref{eq:Rabi_T}) gives $T^{(21)} = 1013$~fs, a value in very good agreement with the simulated full-bath recurrence time. The thus established correspondence with a simple Rabi oscillation between two states supports the idea that the overall dynamics is dominated by a single resonant transition.

However, as highlighted in Fig.~\ref{fig:Dyn_v1_m0}(c), the complete dynamics involves more than these two states, and multiple off-resonant bath states lie close enough in energy to play a non-negligible role. These smaller contributions explain the more sudden changes in the time derivative of the populations, seen typically near 185, 930, 1115 and 1300~fs. The bath frequencies being defined as $\omega_k = \omega_0 + k \times\Delta \omega$, the bath modes closest to resonance are modes $k=20$ and $k=22$ and are associated to the energies $E_{\rm res} \mp \hbar\Delta \omega$. Due to their detuning, these contributions have a smaller amplitude [see Eq.~(\ref{eq:Rabi_P})] and account for 3\% of the total bath population at most. They also have a much shorter period of about 185~fs, in good agreement with Eq.~(\ref{eq:Rabi_T}). Both modes have very similar periods due to their similar coupling strength $V_k$ and their identical detuning $\Delta \omega_k$. Smaller contributions from off-resonant modes give rise to interferences with the main oscillation, leading to the aforementioned marked changes in the variations of the populations. One example can be seen between 900 and 1150~fs in Fig.~\ref{fig:Dyn_v1_m0}(c), where the population in $E_{\rm res}$ drops to its minimum while both states at $E_{\rm res} \pm \hbar\Delta \omega$ almost simultaneously gain in population. The sum of the two populations results in the features noted around 930 and 1115~fs. We also note that, despite seeming sharp and almost non-differentiable, these variations are actually always smooth, as emphasized in the inset of Fig.~\ref{fig:Dyn_v1_m0}. 
When bath frequencies lie even further away from the resonance condition, $\hbar\omega_k = E_{\rm res} \pm n\times\hbar\Delta \omega$ with increasing $n$, their maximum population and their period both rapidly decrease and only the modes $k =$ 20, 21 and 22 keep a maximum population above 1\%.

The dynamics obtained for the same non-resonant bath as studied by Bouakline {\em et al.} in Ref.~\onlinecite{Morse_Surface_2012} is discussed in Appendix \ref{AppB}, and once again the predictions of the EBS model are found to be in quantitative agreement with the MCTDH calculations of these authors.

\subsection{Role of the bath energy grain}
\label{subsec:Delta_E}

The results shown in Fig.~\ref{fig:Dyn_v1_m0} were obtained using a bath energy grain $\Delta E = 2$ cm$^{-1}$.
This parameter directly influences the rounding procedure of the bath frequencies and determines the precision with which the bath energy $m\Delta E$ is discretized. If $\Delta E$ is too large, then the rounded bath frequencies may become excessively approximate, leading to an increase in the detuning between the system and bath frequencies, and potentially causing some essential resonant processes to be ignored.
Conversely, if $\Delta E$ is too small, the numerical cost may increase as a larger number $M$ of effective states becomes necessary to reach the upper limit $E_{\rm max}=M\Delta E$ of the bath energy required to converge the calculations for a given initial state. Hence the value of $\Delta E$ has to be adapted to the system under investigation and the specific initial conditions in order to satisfactorily converge the calculations while keeping a limited numerical cost. 

\begin{figure}
	\includegraphics[width=8.5cm]{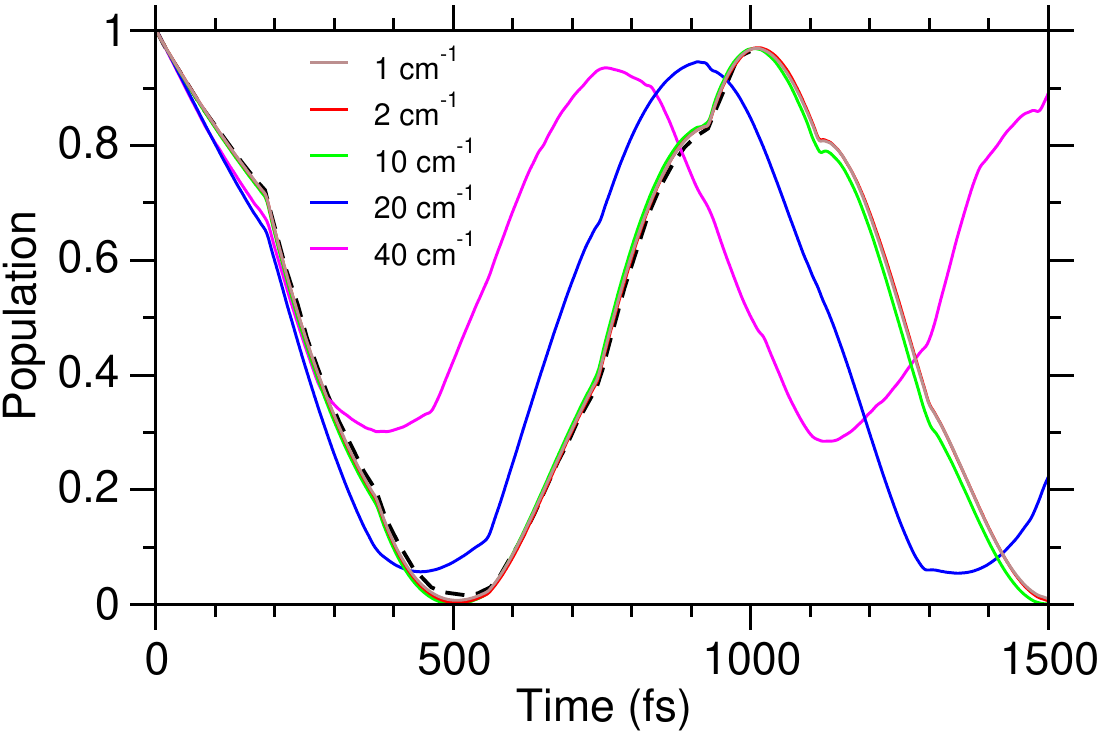}
	\caption{\label{fig:CV_DE} Convergence of the time evolution of the population in the first vibrational excited state with respect to the bath energy grain $\Delta E$ taken in the range 1--40~cm$^{-1}$. The  MCTDH results from Ref.~\onlinecite{Morse_Surface_2012} are superimposed as a dashed line.}
\end{figure}

The quality of the numerical convergence can be estimated from the time variations of the population $P_{v = 1}(t)$ in state $v=1$ for trajectories computed with increasing values of $\Delta E$ at fixed maximum bath energy $E_{\rm max}$. These variations, shown in Fig.~\ref{fig:CV_DE} for $\Delta E$ ranging from 1 to 40~cm$^{-1}$, exhibit no visible difference between the results obtained with $\Delta E= 2$ cm$^{-1}$ and  $\Delta E = 1$ cm$^{-1}$ that both agree very well with the reference MCTDH calculations.
For values of $\Delta E$ above 10~cm$^{-1}$, the results start to differ qualitatively from the reference calculations, with the recurrence time decreasing and the system-bath population transfer becoming less efficient. Considering Eqs.~(\ref{eq:Rabi_T}) and (\ref{eq:Rabi_P}), the decrease in the oscillation time and amplitude are expected to originate from a detuning effect: as the energy grain increases, the rounding of the bath frequencies to the nearest multiple of $\Delta E$ drives $\omega_{21}$ away from $\omega_{0 \rightarrow 1}$, which is not rounded as it is not part of the bath. The value of $E_{\rm res} = m_{\rm res}\Delta E$ is also modified when $\Delta E$ increases and due to these two alterations the system-bath interaction becomes increasingly off-resonant.

Based on these results, and unless stated otherwise, $\Delta E = 2$ cm$^{-1}$ will be used for the calculations below. With this energy grain,  $M = 6\, 000$ effective bath states are typically needed to obtain a converged 2~ps long trajectory starting from $\ket{v_0 = 1,m_0 = 0}$. These parameters yield a maximal internal bath energy of $E_{{\rm{\max}}} \simeq 1.5$ eV, which is about 3.2 times the initial system-bath energy of $\hbar \omega_{0 \rightarrow 1} = 3784.5$ cm$^{-1}$. To study the relaxation of the first vibrational excited state, $N_v = 5$ states are needed for the system. 
If the entire bath had been treated in a similar way, by considering  $N = 5$ states for each bath mode, its full dimensional description would have required $N^g = 5^{40}\approx 10^{28}$ microstates instead of the $M = 6\, 000$ effective energy states used in the present model.

\subsection{Effects of the bath energy}
\label{subsec:Bath_energy}

\begin{figure}
	\includegraphics[width=8.5cm]{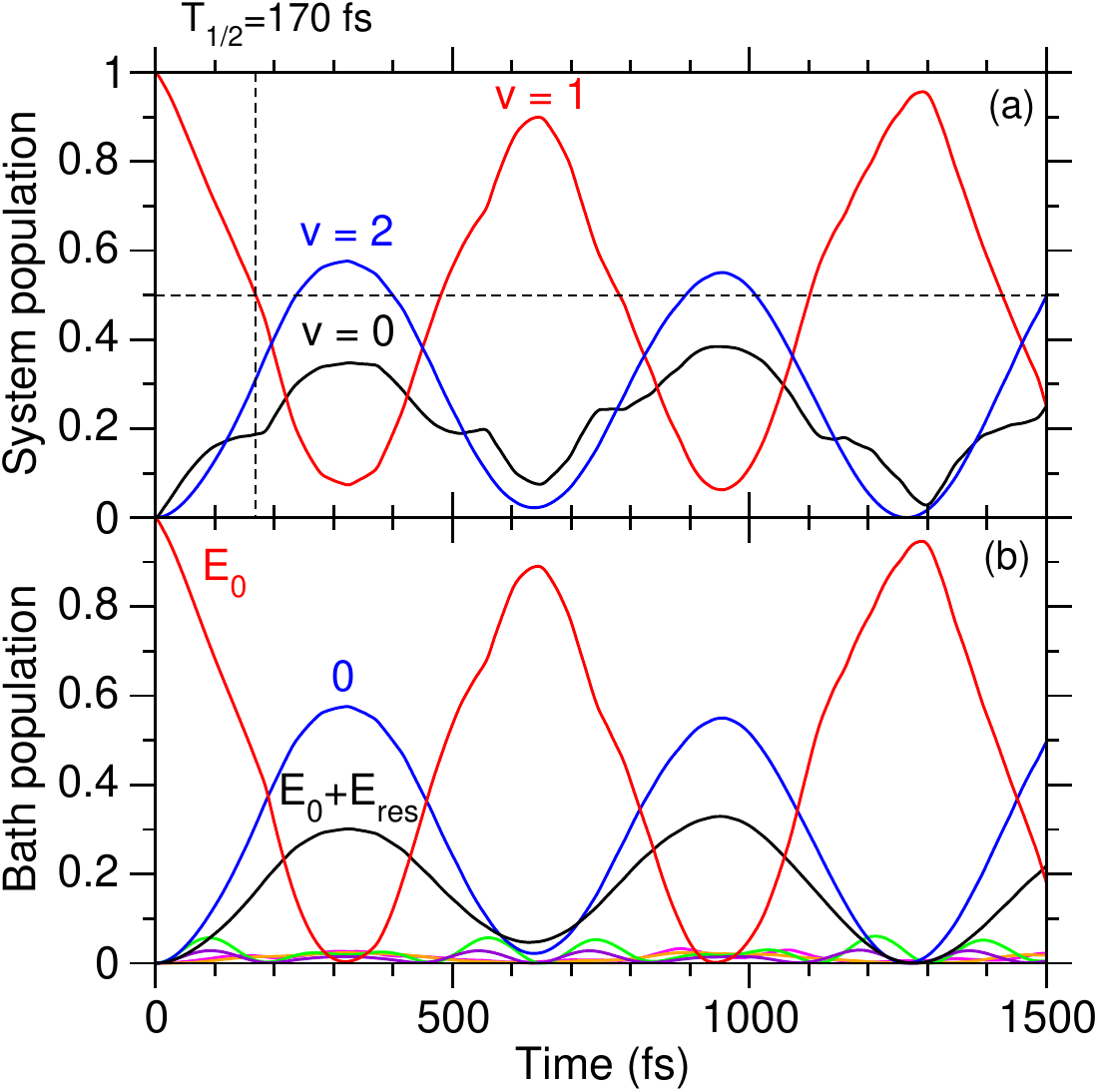}
	\caption{\label{fig:Dyn_v1_m3604}
 Relaxation of a Morse oscillator starting in its first exited state ($v_0 = 1$) and interacting with a bath having an initial energy $E_0 = 3604$~cm$^{-1}=\hbar \omega_{1\rightarrow 2}$. (a) Time evolution of the population in the Morse states $v$; (b) Time evolution of the population of the effective bath states, labeled based on their energy. Smaller contributions from some off-resonant effective states are represented but not labeled for clarity. The energy $E_{\rm res}$ corresponds to the effective state that is resonant with the $v = 1 \rightarrow v = 0$ transition. The time $T_{1/2}$ at which the system population in $v = 1$ drops below 50\% is highlighted by dashed lines.
}
\end{figure}

Within the EBS approach, the bath can be straightforwardly initiated with a finite internal energy. 
Here, the Morse system is initialized in its first excited state $v_0 = 1$, with the bath in a given effective state $m_0 > 0$ associated to a nonzero energy $E_0 =m_0\Delta E$.  Fig.~\ref{fig:Dyn_v1_m3604} shows the time evolution of the system and bath populations obtained for an initial bath energy of $E_0 = 3604$~cm$^{-1}$, which corresponds to the energy of the transition $v = 1 \rightarrow v = 2$.
As can be seen in Fig.~\ref{fig:Dyn_v1_m3604}(a), the introduction of some initial energy into the bath gives rise to a competition between the system transitioning towards $v = 2$ and $v= 0$ as two resonant paths are now accessible to the system from its initial state $v_0 = 1$. While the transition to $v=0$ gives some energy $E_{\rm res} = \hbar \omega_{0 \rightarrow 1}$ to the bath, the transition to $v=2$ retrieves $E_0 = \hbar \omega_{1 \rightarrow 2}$ from it. The maxima reached in the populations of states $v=0$ and $v=2$ of 35\% and 58\%, respectively, are in the same ratio of 1.4 as the corresponding coupling elements $|\bra{v'}f(\hat{z})\ket{v}|$.

As in the case of $m_0 = 0$, some off-resonant transitions also take place and their interferences with the main oscillations cause the sharper variations seen {\em e.g.} in the evolution of $P_{v=0}(t)$.
Adding another channel to empty $v_0 = 1$ also strongly decreases the recurrence time, which drops from 1012 fs to 645 fs. Since these two transitions are not perfectly dynamically synchronized, the system-bath energy transfer is incomplete with 8\% of the population remaining in $v = 0$ and 2\% in $v=2$ at the recurrence time.

\begin{figure}
	\includegraphics[width=8.5cm]{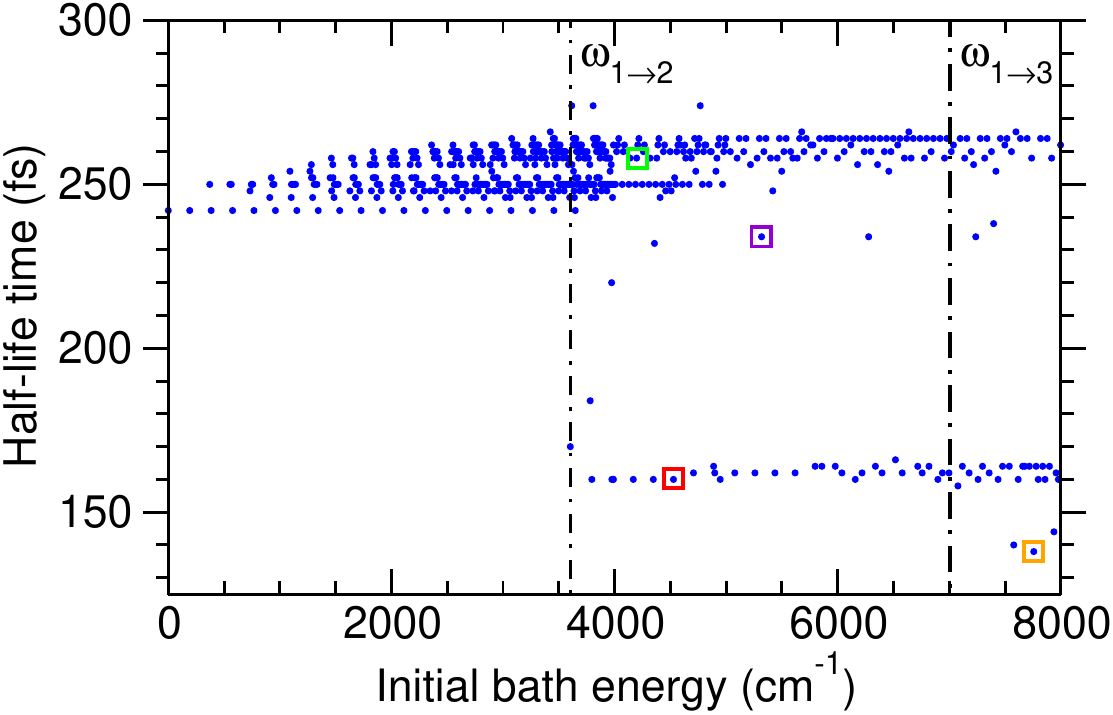}
	\caption{\label{fig:lifetimes_vs_m} Half-life time of $v_0 = 1$ as a function of the initial bath energy. Each blue dot corresponds to a simulation with an initial state $\ket{v_0 = 1, m_0}$ such that $E_0 = m_0\Delta E$ corresponds to the abscissa of the point. Vertical lines indicate the energies of transitions $v = 1 \rightarrow v = 2$ and  $v = 1 \rightarrow v = 3$ respectively. Four points have been highlighted by colored rectangles, the short-time dynamics of the corresponding simulations are shown in Figure \ref{fig:dyn_4_lifetimes}.} 
\end{figure}

The interconversion dynamics between the bath and the system initiated in its first excited state can be further quantified by defining the half-life time $T_{1/2}$ as the time needed to empty half of the initial state population, that is
the shortest time such that $P_{v_0}(T_{1/2})<1/2$. 
In the example of Fig.~\ref{fig:Dyn_v1_m3604}, $T_{1/2}=170$~fs. 
The half-times obtained upon solving the quantum dynamical equations for more than 650 independent bath energies covering the range of 0--8000~cm$^{-1}$ are represented in Fig.~\ref{fig:lifetimes_vs_m}.
In the limit of vanishing bath energy, the half-time amounts to 242~fs. From Fig.~\ref{fig:lifetimes_vs_m}, three regimes can be distinguished: $E_0 < \hbar \omega_{1 \rightarrow 2}$, $\hbar \omega_{1 \rightarrow 2}\leq E_0 <\hbar \omega_{1 \rightarrow 3}$ and $E_0 \geq \hbar\omega_{1\rightarrow 3}$.

If $E_0 < \hbar \omega_{1 \rightarrow 2}$ there is no significant change in the half-life time, which varies between 242 fs and 260 fs. The occasional increase of $T_{1/2}$ is due to the variations in the number of off-resonant transitions that are allowed by the selection rule $\Delta n_k = \pm 1$ and are significant enough to influence the evolution of the initial state. For some initial energies the bath has access to more off-resonant ways to give back its energy to the system, and the system's population that had relaxed to its ground state has a higher probability to be excited again. The rapid re-injection of population leads to an overall slowdown in the decrease of $P_{v=1}(t)$ and to an increase of $T_{1/2}$. 

If $\hbar \omega_{1 \rightarrow 2}\leq E_0 < \hbar \omega_{1 \rightarrow 3}$, the relaxation pathways to $v = 2$ and $v= 0$ can compete with each other. However, this situation does not always occur for arbitrary values of $E_0$ in this range, as the linearity of the coupling in the bath coordinates strongly narrows down the number of possible transitions by allowing only one quantum of energy to be exchanged between the system and one specific bath mode (labeled as mode $k$ in Sec.~\ref{sec:Theory}). Depending on the possible existence of a resonant path from $v = 1$ to $v = 2$, several sub-cases may occur:
\begin{itemize}
    \item [(i)] If the initial bath energy is of the form $E_0 = \hbar \omega_{1 \leftrightarrow 2} + \sum_k n_k\hbar \omega_k$, then it allows for a one-quantum resonant transition towards $v=2$. In this case, both the ground and the second excited states resonantly drain the initial state and the half-life time decreases significantly. This typically corresponds to the situation in Fig.~\ref{fig:Dyn_v1_m3604}, in which $E_0 = \hbar \omega_{1 \rightarrow 2}$ allows for a resonant transition between $\ket{v_0 = 1, E_0 = \hbar \omega_{1 \rightarrow 2}}$ and $\ket{v = 2, E = 0 }$. 
\begin{figure}
	\includegraphics[width=8.5cm]{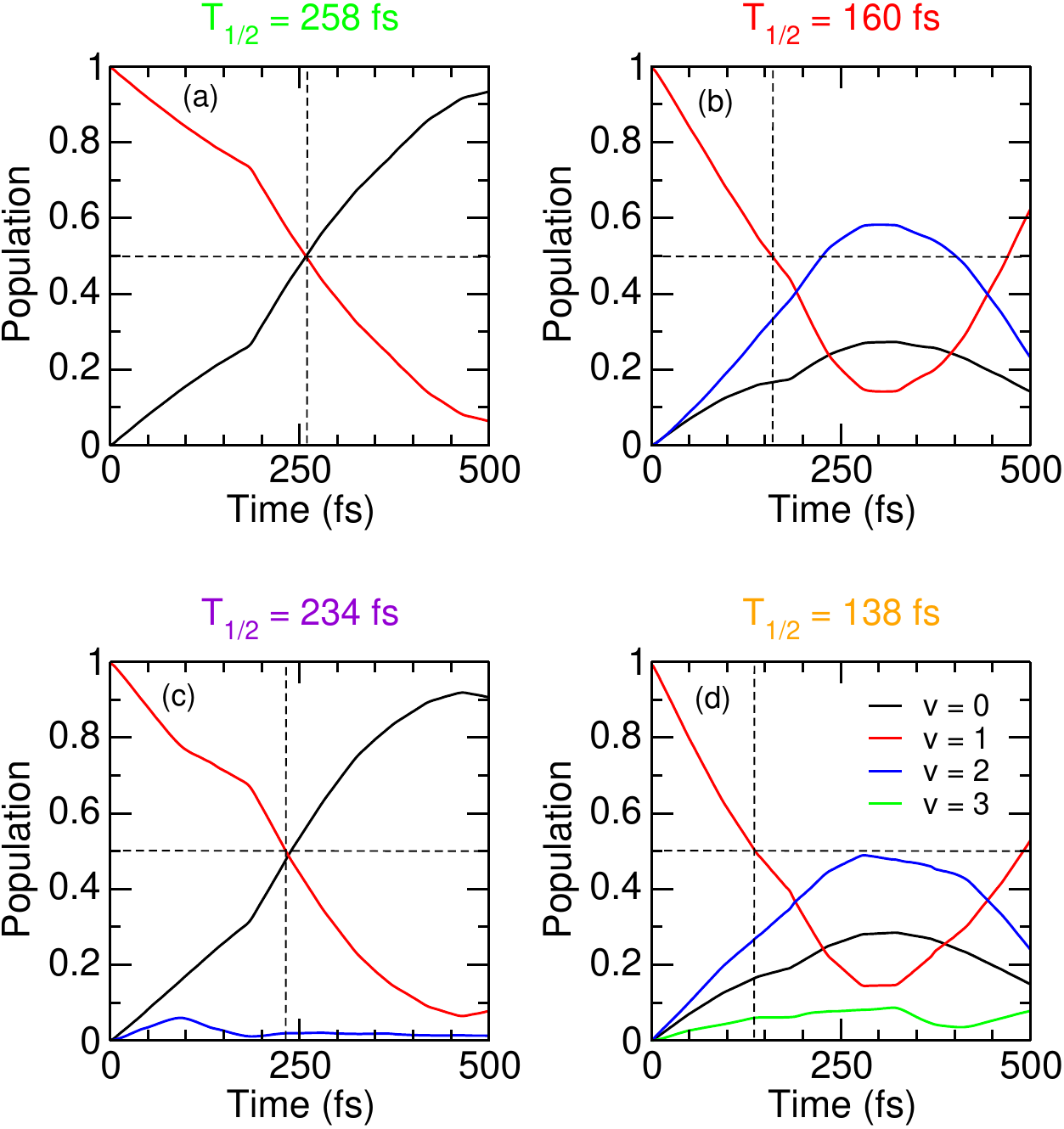}
	\caption{\label{fig:dyn_4_lifetimes} Relaxation of a Morse oscillator starting in its first exited state ($v_0 = 1$) for increasing initial bath energies: (a) $E_0 = 4198$ cm$^{-1}$, (b) $E_0 = 4528$ cm$^{-1}$, (c) $E_0 = 5318$ cm$^{-1}$ and (d) $E_0 = 7758$ cm$^{-1}$. The half-life time associated to each initial condition is indicated in a color referring to Fig.~\ref{fig:lifetimes_vs_m}. Dashed lines emphasize the position of $T_{1/2}$ in each plot.}
\end{figure}
Fig.~\ref{fig:dyn_4_lifetimes}(b) offers another example of such situation, with an initial bath energy $E_0 = \hbar \omega_{1 \rightarrow 2} + \omega_1 + \omega_4$ allowing for a resonant transition from $\ket{v_0 = 1, E_0 = \hbar \omega_{1 \rightarrow 2} + \omega_1 + \omega_4}$ to  $\ket{v = 2, E =\omega_1 + \omega_4}$. In both cases, $v = 2$ gains a significant part of the population, thus accelerating the population decrease in $v = 1$ and leading to half-life times of 170 fs and 160 fs, respectively;

\item [(ii)] If the initial bath energy is not of the form $E_0 = \hbar \omega_{1 \leftrightarrow 2} + \sum_k n_k\hbar \omega_k$, the second excited state will not become significantly populated. Such a situation, illustrated in Fig.~\ref{fig:dyn_4_lifetimes}(a), leads to population evolutions similar to those found for energies lower than $\hbar \omega_{1 \rightarrow 2}$ with a half-life time between 250 fs and 260 fs. Nevertheless, some off-resonant transitions towards $v = 2$ might still be possible, leading to some minor population in the second excited state [see Fig.~\ref{fig:dyn_4_lifetimes}(c)]. In this situation $T_{1/2}$ drops more or less drastically depending on the detuning of the considered transitions with respect to the resonance.
\end{itemize}

If $E_0 \geq \hbar\omega_{1\rightarrow 3}$ then the third excited state might be populated as well, leading to another decrease of $T_{1/2}$ as a third channel contributes to emptying the initial state. As in the previous case, this new channel is effectively open only when a resonant transition $v = 1 \rightarrow v = 3$ is allowed. However, due to the preference of the system for $\Delta v =\pm 1$, this transition is slower than for $v = 1\to v = 2$, and the third excited state hardly gets any population, even in a resonant case such as the one illustrated in Fig.~\ref{fig:dyn_4_lifetimes}(d). The effect on the half-time is thus smaller than in the previously considered cases, with $T_{1/2}$ dropping from 160 fs to around 140 fs.

At thermal equilibrium, canonical properties $A(T)$ at a fixed temperature $T$ can be recovered from the energy-resolved, microcanonical properties $A(E)$ by appropriate statistical reweighting,
\begin{align}
\label{eq:thermal_p}
A(T) &= \frac{1}{Z(T)}\int A(E)\rho(E)\exp(-E/k_{\rm B}T) dE, \\
Z(T) &= \int \rho(E)\exp(-E/k_{\rm B}T) dE,\nonumber 
\end{align}
in which $k_{\rm B}$ and $Z(T)$ denote the Boltzmann constant and the canonical partition function, respectively. This relation can be applied to the time-dependent evolution of the population in the Morse oscillator, placed at time $t=0$ in contact with a thermalized bath and assumed to be in a prescribed state, and in particular to the property $A=P_{v_0}$. Here it should be emphasized that such a canonical averaging procedure only applies to the initial conditions, the subsequent dynamics remaining microcanonical, with the system and bath not being in contact with the thermostat any longer for $t > 0$.
For the same situation covered in Fig.~\ref{fig:dyn_4_lifetimes}, with the Morse system initiated in its first excited state, and considering the rather high frequency of this Morse oscillator,  bath temperatures in excess of 1500~K are required to heat the system even further and enable excitations towards $v >1$, which would probably be an unrealistic temperature regime for the underlying physical system. As a proof of principle, Appendix \ref{AppC} discusses the results obtained under such a situation but for canonical baths at $T=100$--700~K.

\subsection{Effects of the bath size}
\label{subsec:Bath_size}
 
Finally, the capability of the present model to treat large dimensional baths at a reasonable numerical cost is investigated by increasing the number of bath modes. For a given bath size $g$, the bath frequencies are now defined as
\begin{equation}
\omega^{(g)}_k = \omega_0^{(g)} \,+\, k \times\Delta \omega^{(g)}, \quad \forall\; k = 1,\hdots,g \quad,
\label{eq:freq_g}
\end{equation}
with $\Delta \omega^{(g)} =40/g\times\Delta\omega $ and $\omega_0^{(g)}$ such that $\omega^{(g)}_1 = 193.5$ cm$^{-1}$ for all values of $g$. This definition ensures that the frequency range of the bath is preserved by decreasing the frequency gap appropriately, the minimum and resonant frequencies keeping the values of 193.5 and 3782 cm$^{-1}$, respectively. The maximum frequency does not vary significantly and remains in the 7200--7400~cm$^{-1}$ range.
The coupling constants $c_k$ are also adapted by defining 
\begin{equation}
c_k^{(g)} = \omega_k^{(g)} \sqrt{\frac{2\mu_k\mu\gamma\Delta\omega^{(g)}}{\pi}}.
\label{eq:ck_g}
\end{equation}
As the bath size increases, the frequency distribution becomes increasingly dense and more frequencies become near-resonant with the $ v= 1 \rightarrow v = 0$ transition, leading to an exponential decay of both system energy and population into the bath as it tends toward a continuum. 

\begin{figure}
	\includegraphics[width=8.5cm]{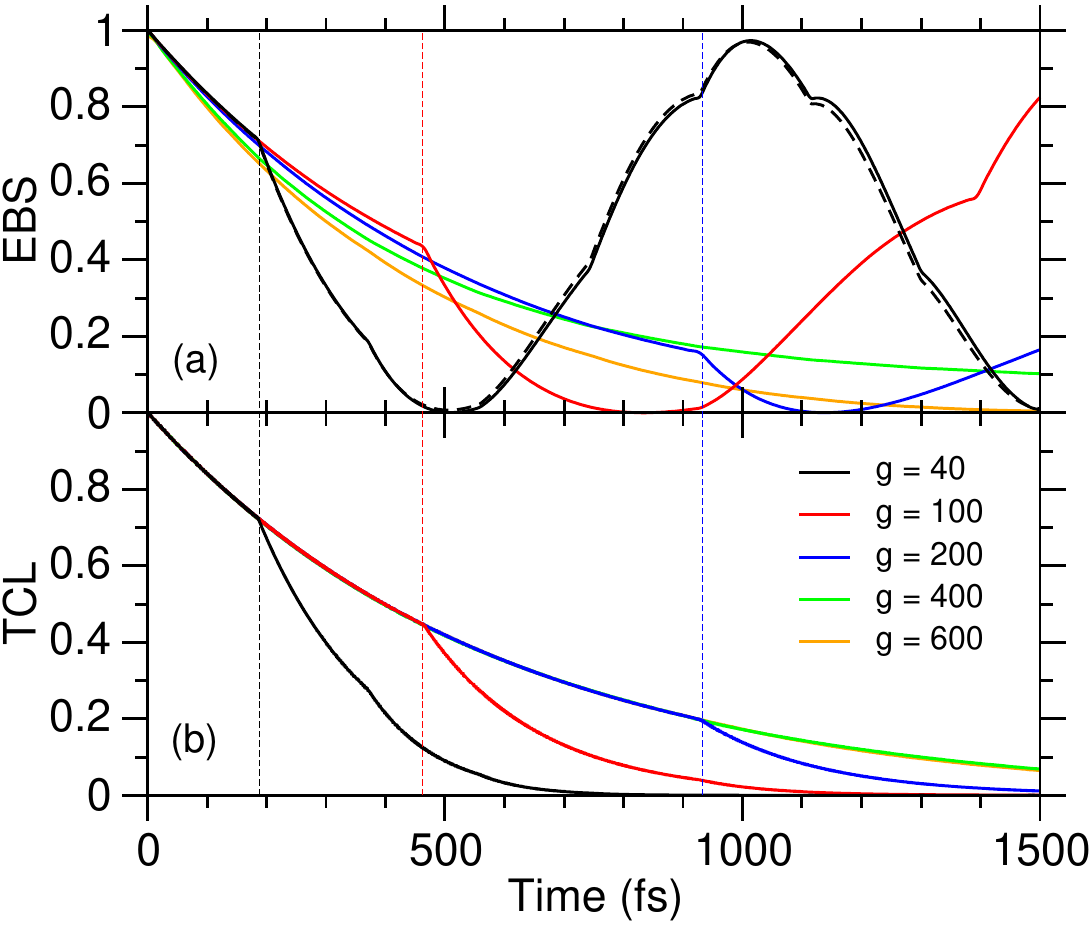}
	\caption{\label{fig:bath_size} Time evolution of the population in  $v = 1$ for different bath sizes $g$, as obtained from (a) the present EBS model, restricting the Morse system to its lowest two states. The dashed curve shows the corresponding results obtained for $g=40$ without such restriction; (b) a second-order time-convolutionless approach (TCL) for the same system. The vertical dashed lines emphasize the occurrence of the sharper variations in the populations, which coincide with both methods.}
\end{figure}

The accuracy of the EBS model for large bath sizes can be assessed by comparing its predictions to those obtained with the second-order time-convolutionless method.\cite{Chang:1993tu,Breuer:2001aa,Breuer:2007ab}  Within this approach, the system dynamics is described by a quantum master equation obtained using the traditional Nakajima–Zwanzig projection operator, in which the bath dynamics is integrated out with a second-order perturbative expansion in the system-bath coupling.  
To simplify the discussion and calculations in this part, without altering the generality of the physical results, only the two lowest levels of the Morse oscillator were considered.
All EBS simulations presented hereafter were performed with $N_v = 2 ,\; M = 24\,000,$ and $\Delta E = 0.5$ cm$^{-1}$. The population in the first excited state $v_0 = 1$ obtained using the EBS model initiated with the bath in its ground state is represented in Fig.~\ref{fig:bath_size}(a) as a function of time, for different bath sizes.
For $g=40$, the results obtained with the two-state model are very close to those from the complete Morse model, indicating that excited states $v > 1$ have a minor contribution, which further justifies this approximation for the present discussion. Fig.~\ref{fig:bath_size}(b) shows the corresponding results obtained with the TCL approximation, for the same two-level system and bath sizes extending to $g=600$.

From Fig.~\ref{fig:bath_size}(a), the population in the first excited state decreases exponentially up to a certain time $\tau$ that increases with the bath dimension, before suddenly dropping to zero. This behavior is quantitatively reproduced by the TCL method in Fig.~\ref{fig:bath_size}(b). At longer times, recurrences take place due to the conservation of total energy in the microcanonical ensemble, and the bath converts a part of its energy back to the system. Such recurrences are only found with the EBS method and are necessarily missing from the TCL calculation which assumes that the bath is at thermal equilibrium and only soaks up the energy from the system. 

The decay in the system population at short times predicted by the TCL approach is thus well reproduced by the present EBS approach, minor differences being noticeable only for the largest bath size of $g = 600$. Such discrepancies are numerical in nature and probably arise from the strong irregularities and very high values reached by the density of states in a high-dimensional, perfectly harmonic system. In addition, as $g$ increases from 40 to 600,
the gap $\Delta \omega^{(g)}$ between successive bath frequencies decreases from 180~cm$^{-1}$ down to 12~cm$^{-1}$ and becomes comparable to $\Delta E$, reaching a regime in which the approximations underlying the EBS method become more questionable.

The discussion above shows that the EBS method is particularly useful for finite but large baths, for which full-dimensional wavefunction-based methods are impractical. Instead of propagating a full dimensional wavefunction that would require a cumbersome basis set $\ket{v}\otimes\ket{n_1}\otimes \ket{n_2}\otimes ...\otimes \ket{n_{g}}$ of size $\sim N_v\times N^g$, the present EBS model uses the eigenstates of the effective Hamiltonian, which are in much lower number $N_v\times M$. 
This raises the question of its computational cost, which we discuss briefly here before concluding. With the present parameters, calculations on Intel E5-2670 V3 processors and parallelized on eight CPUs take 70 and 170 minutes for baths of dimensions 100 and 600, respectively, further indicating a rather favorable scaling. This computational cost is almost entirely contained in the diagonalization step of the effective Hamiltonian and is thus mostly sensitive to the sparseness of the matrix.

\section{Conclusion and perspectives}
\label{sec:conclusion}

In this work, a new theoretical approach to model the quantum dynamics of a one-dimensional anharmonic system interacting with a finite but large dimensional harmonic bath was developed. In our effective bath states model, the bath is coarse-grained into effective energy states that are defined based on their total energy, which is discretized using an energy grain $\Delta E$.
These effective energy states operate through rigorous coupling with the system that accounts for the number of accessible microstates within the bins involved with each transition. The coarse-graining procedure implicitly assumes that all microstates within an effective energy state are equiprobable and satisfy microcanonical statistics.

This EBS approach strongly reduces the effective dimension of the bath and its increase with the number of bath modes, while keeping some information about the global state of the bath and its evolution with time. The bath energy grain is treated as a parameter that has to be adjusted depending on the desired balance between accuracy and computational time for a given problem. As an example, to study the vibrational relaxation of a molecule $\Delta E$ would typically be of the order of 1~cm$^{-1}$.
The interactions between the system and individual bath modes are included in the couplings and all the bath microstates falling into a prescribed energy grain are taken into account through the identification of active and spectator bath modes and the calculation of various densities of states to construct the required microcanonical probabilities.
While a linear coupling in the bath coordinates was considered in our first application, the EBS model can deal with any polynomial coupling as well, thus allowing more realistic cases to be considered. For example, a quartic potential energy surface obtained from quantum chemistry calculations within vibrational perturbation theory could be used to parametrize a polynomial system-bath coupling Hamiltonian and to model, \textit{e.g.} intramolecular vibrational redistribution in a molecular system.

The strategy of the present EBS method relies on the use of the system eigenstates and the bath effective energy states to construct an effective Hamiltonian and obtain its eigenelements, from which the time-dependent Schrödinger equation can be solved exactly, allowing for long time scales to be considered even for large dimensional baths. 

The model was validated by comparison with earlier calculations based on the multiconfigurational time-dependent Hartree method, for a Morse oscillator coupled to 40 harmonic oscillators.\cite{Morse_Surface_2012} In both resonant and non-resonant cases, the relaxation dynamics following preparation of the system onto its first excited state, with the bath in its ground state, was found to be in very good agreement with the MCDTH results provided that the energy grain is small enough. The dimensionality reduction brought by coarse-graining the bath allowed us to consider much larger --- but still finite --- baths amounting to 600 oscillators at a very reasonable computational cost. In such cases, the energy flow dynamics from the system to the bath predicted by the EBS method was found to be in quantitative agreement with the results of the more approximate time-convolutionless method, the time at which the energy accumulated in the bath starts flowing back to the system --- a feature that the TCL approach cannot describe --- increasing significantly with the bath size.

More realistic cases where the bath is prepared at a finite energy were also considered. Under such situations, the quantum dynamics becomes more involved as new relaxation channels are opened, for the system and the bath alike. One particularly valuable feature of the EBS approach is that it allows for such microcanonical simulations to be performed routinely, making the finite temperature preparation of the bath also accessible, canonical statistics being accounted for by appropriate reweighting of energy-resolved (microcanonical) datasets. 

The EBS model is naturally suited to address the intramolecular quantum dynamics of large, isolated molecules. In particular, intramolecular vibrational relaxation (IVR) processes occurring subsequently to the infrared excitation of a particular mode, or following the optical excitation of a larger chromophore and its internal conversion, can be investigated in detail, even for molecules containing tens or even a few hundred of atoms. The model could also be extended to treat microsolvated species and describe the early stages of energy redistribution between the solute and the solvent molecules. Among the valuable features of our approach, vibrational spectra of a mode of interest (the system) influenced by the rest of the molecule (the bath) can be determined at fixed temperature and monitored as a function of time, as in pump-probe experiments. In practice, these applications will require developments to account for multidimensional systems, including anharmonic baths as well as more complex system-bath coupling terms along the lines suggested above.

Finally, one possible field of application of the current approach is that of synchronization dynamics of quantum oscillators. This field has attracted significant interest in recent years,\cite{lorchanb16,benedettigmz16} in particular to clarify the relation between synchronization and entanglement in quantum oscillators\cite{rouletb18,manju23} and the role of the environment.\cite{henriet19} Here it should be possible to consider a reduced number of oscillating systems, coupled to each other directly and through their interaction with a finite, but possibly large-dimensional bath, and to investigate their dynamics over long times.

\begin{acknowledgments}
The authors wish to thank Dr. F. Bouakline for useful discussions and financial support from GDR EMIE 3533. L.A. acknowledges financial support from MESRI through a PhD fellowship granted by the EDOM doctoral school.
\end{acknowledgments}






\section*{Data Availability Statement}
The data that support the findings of this study are available from the corresponding author upon reasonable request.

\appendix

\section{Dynamics of a non-resonant bath}
\label{AppB}

In this appendix we consider the case of the non-resonant bath already investigated by Bouakline {\em et al.}\cite{Morse_Surface_2012} in their original study of the same model. The non-resonant bath is also defined by a linear distribution of the bath frequencies but with $\omega_0 = 0$ and $\Delta \omega =153.6$~cm$^{-1}$. It is  non-resonant with the Morse system, as none of the bath modes defined in this way are resonant with the $v = 1 \rightarrow v = 0$ transition. As in Sec.~\ref{subsec:v1_m0} the dynamics with the EBS model was initiated from $\ket{v_0 = 1, m_0 = 0}$.

\begin{figure}[t]
	\includegraphics[width=8.5cm]{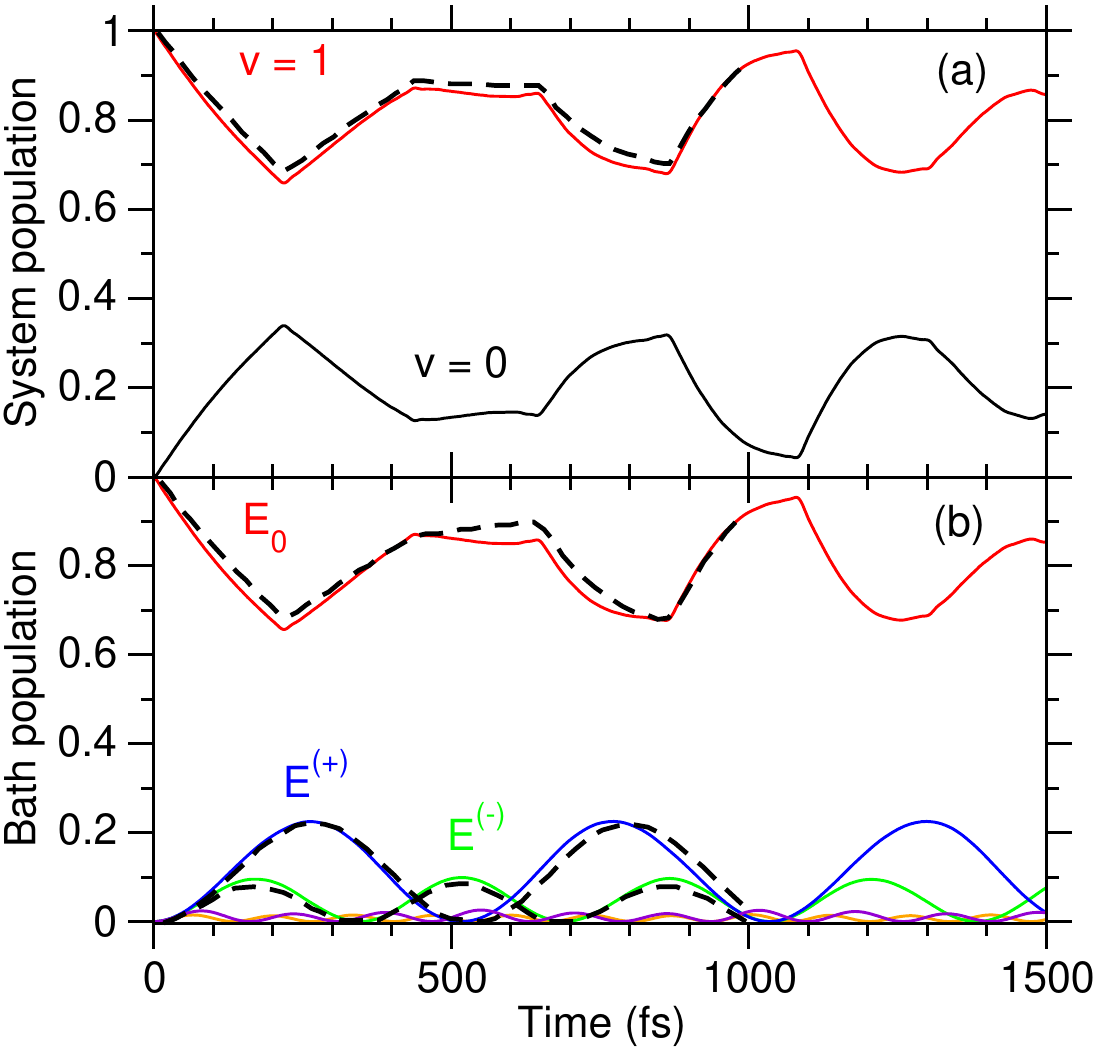}
	\caption{\label{fig:Non_resonant_bath}  Time evolution of a Morse oscillator starting from its first exited state ($v_0 = 1$) interacting with a non-resonant bath of 40 harmonic oscillators initially in its ground state ($E_0 = 0$). (a) Time evolution of the population in the Morse states $v$. (b) Time evolution of the population of the effective bath states, labeled by their energy. The energy $E^{(+)}$ (resp $E^{(-)}$) corresponds to the closest bath mode above (resp below) the frequency of the $v = 1 \to v = 0$ transition. In both panels, the MCTDH results from Ref.\onlinecite{Morse_Surface_2012} are superimposed as dashed lines.}
\end{figure}

The results are shown in Figure \ref{fig:Non_resonant_bath}, in comparison with the MCTDH results of Ref.~\onlinecite{Morse_Surface_2012}.
In this case, the system mostly interacts with the two closest bath frequencies at $\omega^{(+)} = 3840$~cm$^{-1}$ and $\omega^{(-)} = 3686$~cm$^{-1}$, out of tune by $\Delta \omega^{(+)} = 55.5$~cm$^{-1}$ and  $\Delta \omega^{(-)} =98.5$~cm$^{-1}$, respectively. These frequencies are associated to effective bath states with energies $E^{(\pm)}$ that produce the bath dynamics as depicted in Figure \ref{fig:Non_resonant_bath}(b).
From Eq.~(\ref{eq:Rabi_T}), the Rabi periods associated to these two modes are $T_+ =526$~fs and  $T_- =324$~fs, respectively. The approximate ratio of 1.6 between $T_+$ and $T_-$ explains the plateau observed between 440 and 650~fs, as the two modes are almost in opposite phase, each of them gaining population as the other is being depleted. At longer times, the dynamics changes as the two modes are not perfectly synchronized, producing smoother features.
The relative amplitudes of the two off-resonant contributions can be explained by the off-resonant Rabi formula of Eq.~(\ref{eq:Rabi_P}), given that the coupling element $V_k$ is mostly the same for both modes but $\omega^{(+)}$ is closer from the resonant condition than $\omega^{(-)}$.

\section{Temperature effects}
\label{AppC}

The EBS model can be applied assuming an initially thermalized bath at fixed temperature $T$, using the results at fixed bath energy $E$ reweighted by the corresponding thermal probability [see Eq.~(\ref{eq:thermal_p})].
Assuming the system to be initially in its first excited state, the set of trajectories discussed in Sec.~\ref{subsec:Bath_energy} can be used to predict the probability that it stays on this level as a function of time, after canonical averaging of the energy-resolved probabilities.

\begin{figure}[t]
	\includegraphics[width=8.5cm]{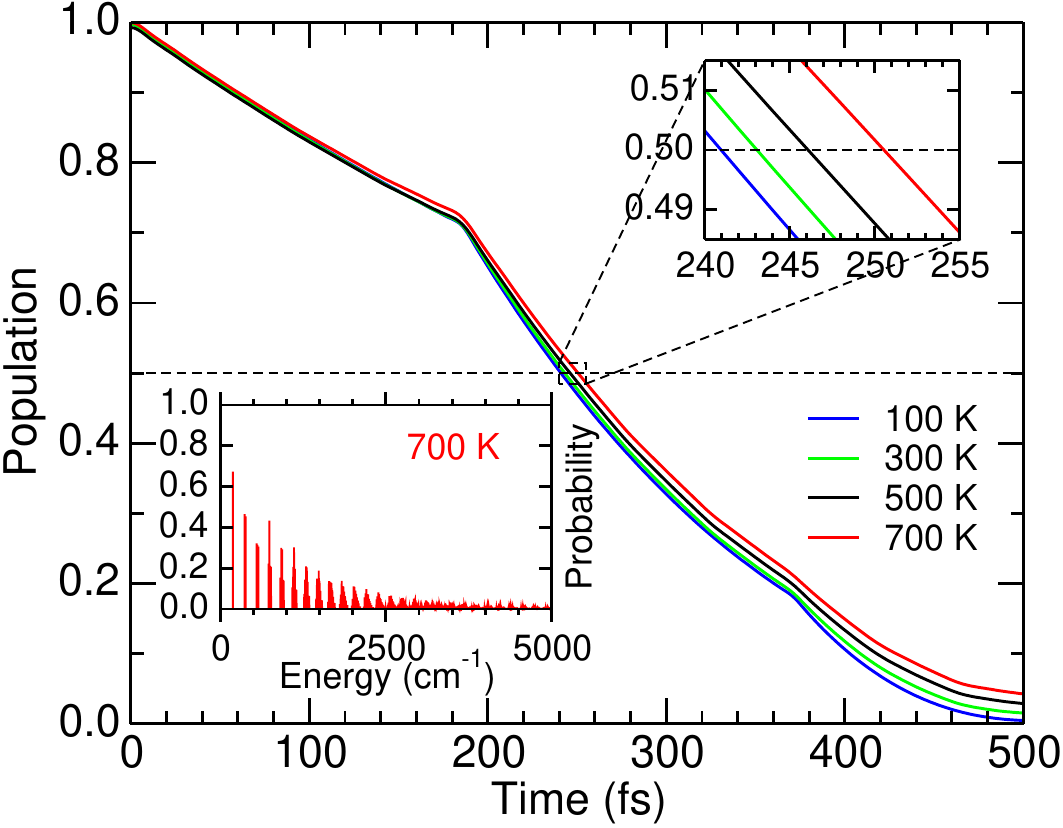}
	\caption{Probability for the system to be in its first excited state, after being placed in contact with a bath described by canonical statistics at temperatures $T=100$, 300, 500, and 700~K. The inset in the upper right corner highlights the time interval where the probability crosses 50\%, while the inset in the lower corner shows the canonical probability of the bath being in a specific energy state, at 700~K.}. \label{fig:T_effect}
\end{figure}
These probabilities obtained at four temperatures in the range 100--700~K are shown in Fig.~\ref{fig:T_effect} as a function of time.
The half-time $T_{1/2}$ is of the same order of magnitude as the values obtained in Fig.~\ref{fig:lifetimes_vs_m} at low energies, and this is consistent with the rather low bath energies obtained at these temperatures from the canonical statistics, which barely extend above 2500~cm$^{-1}$ at 700~K. The half-time slightly increases with increasing temperature, also in agreement with the general trend in Fig.~\ref{fig:lifetimes_vs_m} at low energies. The very low magnitude of this effect (less than 10~fs in the entire temperature range) is the manifestation that even at 700~K the energy contained in the bath is not sufficient to significantly populate the excited states of the Morse oscillator.

\bibliography{Biblio}

\end{document}